\documentclass[aps,nofootinbib]{revtex4-1}

\usepackage{relsize}
\usepackage[intlimits]{amsmath}
\usepackage{amsmath}
\usepackage{blkarray}
\usepackage{exscale}

\usepackage{graphicx}
\usepackage{dcolumn}
\usepackage{bm}
\usepackage{multirow}                               
\usepackage{subfig}
\usepackage{array,ragged2e}

\usepackage{float}
\usepackage[justification=justified]{caption}
\usepackage{caption}
\usepackage{mathtools}
\usepackage{epstopdf}
\usepackage[toc,page]{appendix}

\usepackage[font={small}]{caption}

\usepackage{multirow}
\usepackage{graphicx}
\usepackage{booktabs}
\usepackage{epstopdf}
\usepackage{color}

\begin{document}


\title{Emittance reduction with variable bending magnet strengths: \\
Analytical optics considerations and application to the Compact Linear Collider damping ring design\\}

\author{S.~Papadopoulou}
\affiliation{European Organization for Nuclear Research (CERN), BE Department, 1211 Geneva 23, Switzerland}
\affiliation{Department of Physics, University of Crete, Greece}
\affiliation{Institute of Theoretical and Computational Physics (ITCP), Crete, Greece}
\author{F.~Antoniou}
\affiliation{European Organization for Nuclear Research (CERN), BE Department, 1211 Geneva 23, Switzerland}
\author{Y.~Papaphilippou}
\affiliation{European Organization for Nuclear Research (CERN), BE Department, 1211 Geneva 23, Switzerland}
\date{\today}

\begin{abstract}
\par  
One of the main challenges of the lattice design of synchrotrons, used as light sources or damping rings (DRs), is the minimization of the emittance. The optimal lattice configurations for achieving the absolute minimum emittance are the theoretical minimum emittance (TME) cells. This paper elaborates the optimization strategy in order to further reduce the betatron emittance of a TME cell by using dipoles whose magnetic field varies longitudinally.  
Based on analytical results, the magnet design for the fabrication of variable bends with the optimal characteristics is discussed.  In order to have a global understanding of all cell properties, an analytical approach for the theoretical minimum emittance (TME) cells with variable bends is elaborated. 
This approach is employed for the design optimization of the Compact Linear Collider (CLIC) DRs. The margin gained in the emittance including IBS based on this new design strategy enables the removal of a number of TME cells from the existing arcs while still keeping the requirements of the collider. The reduction of the circumference is further enhanced by the use of optimized high-field wigglers. The optimization strategy followed for the CLIC DRs is explained in detail and the output parameters of the new design are presented.

\end{abstract}
\maketitle
\section{Introduction}
\par The principle objective of linear collider damping rings (DRs) and storage ring X-ray sources is the generation of ultra-low emittances in order to achieve high brightness beams. The optics design is primarily focused on building a compact ring, attaining a sufficiently low emittance and an adequately large dynamic aperture (DA). Due to their compactness and the very small horizontal emittance they reach, the TME cells~\cite{ref:tme0,ref:tme} are preferred for some ring designs. The performance of these cells was previously studied analytically for the case of uniform dipoles~\cite{ref:fan}. The emittance reached by a TME cell can be further reduced if instead of standard dipoles, longitudinally variable bends~\cite{ref:gr,ref:nw} are used. This paper summarizes analytical and numerical studies regarding the variable bends that provide emittances lower than the ones of a uniform dipole for a TME cell. The optimal magnetic field evolution of these bends
is found analytically and applied to two simple dipole profiles (Section~\ref{dipprof}). 
The characteristics of the dipole with the optimal field variation are further refined by the technological limitations for the magnet fabrication (Section~\ref{magdesign}). These limitations are used as constraints in order to study the impact of each dipole profile on the optics functions of the cell and on the properties of the ring.
The analytical parametrization of the quadrupole strengths and optics functions with respect to the drift lengths and the emittance is extended to the TME with a non-uniform dipole (Section~\ref{vbtme}). 
The strong focusing needed for accomplishing the TME conditions results in cells with high chromaticities. In favor of having low chromaticity, a sufficient large dynamic aperture and a compact ring, the TME should be detuned from the absolute minimum emittance solution. 
\par Based on the analytical thin-lens solutions, appropriate initial conditions are chosen for matching the lattice numerically using MADX~\cite{ref:madx}. This approach is applied to the CLIC DRs, filled with longitudinally variable bends in the arcs and high field wigglers in the long straight sections. The optimization strategy followed to reduce the circumference of the DRs design is explained in Section~\ref{CLICVB}.  To this end, the parameters of the new alternative design are compared with respect to the original ones.  
\section{Analytical solutions for minimizing the emittance of a TME cell}

\par 
The equilibrium horizontal emittance in a storage ring is given by:
\begin{equation}
\label{eq:epsi}
\epsilon_{x}=\dfrac{C_{q}\gamma^2}{J_{x}} \ \dfrac{\langle\dfrac{\mathcal{H} }{\left| \rho \right|^3 }\rangle}{\langle \dfrac {1}{\rho^2}\rangle}=
\dfrac{C_{q}\gamma^2}{J_{x}} \ \dfrac{\dfrac{1}{C} \mathlarger{\int\limits}_0^C \dfrac{\mathcal{H} }{\left| \rho \right|^3 }ds} {\dfrac{1}{C} \mathlarger{\int\limits}_0^C \dfrac{1}{\rho^2}ds}\;\;,
\end{equation}
where $C$ is the circumference of the ring, $\gamma$ is the Lorentz factor, $J_x$ is the damping partition number and $C_{q}=3.84 \times 10^{-13}$ m (for electrons). The lattice function ${\cal H}$ known as the dispersion invariant,  depends on the optics parameters, the dispersion function and its derivative: 
\begin{equation}
\label{eq:H}
{\cal H}(s) = \gamma(s) \eta(s)^2 + 2 \alpha(s) \eta(s) \eta(s)^\prime + \beta(s) {\eta(s)^\prime}^2\;\;.
\end{equation}
In the case of uniform dipoles, having a constant bending radius $\rho$, the minimum emittance value is obtained through the minimization of the $\left\langle {\cal H} \right\rangle$.  However, in the case of longitudinally variable bends, for a varying $\rho$ along the length of the magnet, the aim is to minimize $\left\langle {\cal H}/\rho^3 \right\rangle/\left\langle 1/\rho^2 \right\rangle$. 

\par
\begin{figure}[h]
\centering
\includegraphics[width=0.4\linewidth]{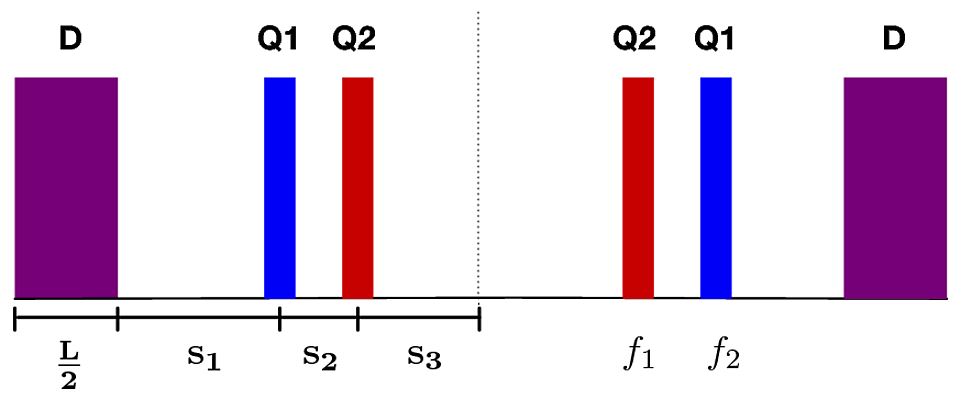}
\caption{\small {Schematic layout of a TME cell}}
\label{fig:tmecell}
\end{figure}

  A schematic layout of the TME cell is displayed in Fig.~\ref{fig:tmecell}. It consists of one dipole D of length $L$ and of two quadrupole families Q1, Q2 with focal lengths $f_1[m] = 1/(k_1l_{q_1})$ and $f_2[m] = 1/(k_2l_{q_2})$, where $k_1,k_2$ are the normalized quadrupole strengths and $l_{q_1}, l_{q_2}$ their lengths. The drifts between the elements are denoted by $s_1$, $s_2$ and $s_3$.
For simplicity, the center of consecutive dipoles is considered as the entrance and
exit of the TME cell, with the index ``cd'' (center of dipole) representing functions at this location. The optics parameters, the dispersion function and its derivative from the center to the edge can be written as:
\begin{equation}
\label{eq:b}
\beta\left( s \right) = \beta_{cd}-2\alpha_{cd} s+ \gamma_{cd}s^2 
,
~
\alpha \left( s \right)=\alpha_{cd}-\gamma_{cd}s
,
~
 \gamma \left( s \right)=\gamma_{cd}
,
~
\eta \left( s \right)=\eta_{cd} +\eta'_{cd}s+  \tilde{\theta} \left( s \right) 
,
~
\eta' \left( s \right)=\eta'_{cd}+\theta\left( s \right)\;\;.
\end{equation}
The expressions in Eq.~\eqref{eq:b} are used to calculate the dispersion invariant ${\cal H}(s)$ (Eq.~\eqref{eq:H}). It should be mentioned that the focusing of the dipole is negligible for dipoles having a small bending angle $\theta$.
\par
\begin{figure}[h]
\centering
\includegraphics[width=0.4\linewidth]{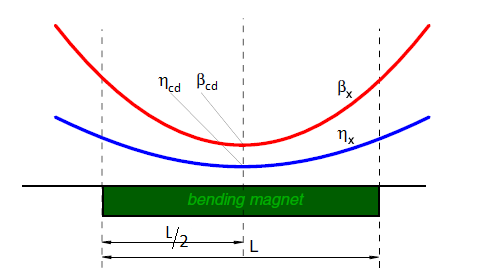}
\caption{{\small Symmetry condition for the TME.}}
\label{fig:tme}
\end{figure}

   The theoretical minimum emittance can be achieved if the symmetry condition, for which both beta ($\beta_x$) and dispersion ($\eta_x$) functions have a minimum at the center of the bending  magnets ($\alpha_{cd}=\eta'_{cd}=0$), is satisfied (Fig.~\ref{fig:tme})~\cite{ref:tme0,ref:tme}.
For isomagnetic TME cells the dispersion and beta functions at the center of the dipole are respectively equal to
$
\eta_{cd}=\frac{\theta L}{24}
$ and
$
\beta_{cd}=\frac{L}{2\sqrt{15}}
$. These functions are different for a non-uniform dipole, since their bending angle and radius vary along the electron beam path in the magnet. In that case, the bending angle and its integral are given by:
\begin{equation}
\label{eq:th}
\theta=\mathlarger{\int\limits}_0^s \dfrac{1}{\rho(s)} ds
~,~~
\tilde{\theta}=\mathlarger{\int\limits}_0^s \left( \mathlarger{\int\limits}_0^s \dfrac{1}{\rho(s)} ds \right) ds\;\;.
\end{equation}

\par
The beta and dispersion functions at the dipole center  ($\beta_{cd}$ and $\eta_{cd}$) impose two independent
optics constraints, therefore, at least two quadrupole families are needed for achieving them.  Using the thin-lens approximation and for given $\beta_{cd}$  and $\eta_{cd}$ at the center of the dipole, the analytical expressions for the quadrupole focal lengths can be derived: 
\begin{equation} 
f_{1}=\dfrac{s_{2}g_1}{g_1-\eta_{ss}   + s_{2}\theta }\;\;,\quad
f_{2}=\dfrac{s_{2} \eta_{ss}}{g_1-\eta_{ss}}\;\;, 
\label{eq:f}
\end{equation}
where the dispersion $\eta_{ss}$ at the center of the cell, between two mirror symmetric quadrupoles, 
\begin{equation}
\label{eq:hss}
\eta_{ss}=\dfrac{\dfrac{-2g_1 s_3}{s_2}}{1 \pm \sqrt{1 + \dfrac{4g_1 s_3}{s_2^2} \dfrac{\beta_{cd}^2 \theta - \left(L/2 + s_1\right)g_2}{\beta_{cd}^2 \theta^2+g_2^2} }}\;\;,
\end{equation}
 depends on the drift lengths, the optics functions at the dipole center, the bending characteristics and the parameters 
\begin{equation}
g_1= \eta_{cd} + s_{1}  \theta + \tilde{\theta},\quad
g_2=\eta_{cd} - \dfrac{L}{2}\theta + \tilde{\theta} \;\;.
\label{eq:g}
\end{equation}


\par In the limit of $s_2 \to 0$, i.e. when Q1 and Q2 are merged into one quadrupole, both $f_1$ and $f_2$ vanish, giving infinite focusing strengths (see Appendix~\ref{focal}). When  $s_3 \to 0$, the two quadrupoles that belong to the Q2 family of the TME cell are merged into one and $f_2$ vanishes. In this respect, ultra-short drift space lengths between the quadrupoles must be avoided. The fact that in the limit of $s_1 \to 0$ both $f_1$ and 
$f_2$ have fixed non-zero values, implies that it is possible to place the quadrupole Q1 exactly next to the dipole, with no drift space between them.  

\par
\begin{figure}[h]
	\centering
	\includegraphics[width=0.32\linewidth]{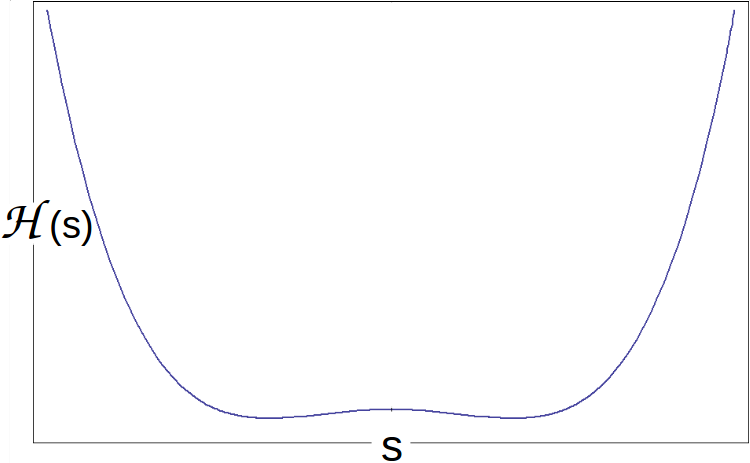}
	\caption{{\small The evolution of the dispersion invariant along the uniform dipole.}}
	\label{fig:Huni}
\end{figure}

 Longitudinally variable dipoles, whose magnetic field varies along their length,  can provide lower horizontal emittances than a uniform dipole of the same bending angle~\cite{ref:gr,ref:nw}. In the case of a TME cell, the evolution of ${\cal H}(s)$ along a uniform dipole is shown in Fig.~\ref{fig:Huni}. 
This evolution guides the bending radius choice for achieving an emittance reduction by using a variable bend. In fact, the variable bend should be designed such that the minimum bending radius is at the dipole's center and then it should decrease towards the edge of the dipole~\cite{ref:pe,ref:esrf,ref:wwp,ref:wang,ref:ler14_stef, ref:streun15,ref:ipac15_stef,ref:esrf2}. 
\par  For the present work, two different functions of the bending radius are used, being either constant or linearly dependent with respect to the distance $s$. Due to the optics symmetry, all definitions can be given for the half-dipole (from 0 to $L/2$).  In this respect, the first part of the half dipole at the dipole center ($s=0$) has a length $L_{1}$ and the second, with a length $L_{2}$, follows until the exit of the dipole, with functions $\rho_1(s)$ and $\rho_2(s)$, describing the bending radii in the respective parts.  
\par The maximum magnetic field is at the center of the dipole, where the bending radius is minimum. The minimum magnetic field and maximum bending radius are at the edges of the magnet. The field evolution along the magnet can be well described by using only two parameters; the lengths and the bending radii ratio~\cite{ref:wang, ref:ler14_stef, ref:ipac15_stef}, that are defined as:
\begin{equation}
\lambda=\dfrac{L_{1}}{L_{2}}~~~\mathrm{and}~~~
\tilde{\rho}=\dfrac{\rho_{1}}{\rho_{2}}\;\;.
\end{equation} 
The lower  $\lambda$ is, the shorter is the high field region compared to the low field one. The lower $\tilde{\rho}$ is, the higher is the field at the dipole center compared to the one at the edges. Generally,  $\tilde{\rho}<1~\mathrm{because }~\rho_2~\mathrm{needs~to~be~larger~than}~\rho_1~\mathrm{and}~\lambda>0$.
\par Splitting the half dipole in two parts with different bending radii  requires the study of the dispersion invariants for each part separately. In this respect, the dispersion invariant given in Eq.~\eqref{eq:H} should be calculated for both parts; that is ${\cal H}_1 (s)$ and ${\cal H}_2 (s)$ with:
\begin{equation}
\label{eq:H12}
{\cal H}_{1,2} (s)= \gamma_{1,2} {\eta_{1,2}}^2 + 2 \alpha_{1,2} \eta_{1,2} \eta'_{1,2} + \beta_{1,2} {\eta'_{1,2}}^2\;\;.
\end{equation}
After implementing the symmetry condition in Eq.~\eqref{eq:b}, the optics functions  for the first and the second part of the half dipole are given by Eq.~\eqref{eq:optics1} and Eq.~\eqref{eq:optics2} respectively, for $\beta_{L_1},~\alpha_{L_1},~\gamma_{L_1},~\eta_{L_1}$ and $\eta'_{L_1}$ corresponding to the optics functions at the point where $s=L_1$.
\begin{equation}
\label{eq:optics1}
\beta_1= \beta_{cd}+\gamma_{cd}s^2  
, ~
\alpha_1=-\gamma_{cd}s 
, ~
\gamma_1=\gamma_{cd}
,~
\eta_1=\eta_{cd}+\tilde{\theta}_1  
,~
\eta'_1=\theta_1 \hspace{0.4cm} (\alpha_{cd}=0,\eta'_{cd}=0)
\end{equation}
\begin{equation}
\begin{aligned}
\label{eq:optics2}
\beta_2= \beta_{L_1}-2(s-L_1)&\alpha_{L_1}+(s-L_1)^2\gamma_{L_1} 
, ~
\alpha_2=\alpha_{L_1}-(s-L_1)\gamma_{L_1}
,~
\gamma_2=\gamma_{L_1}
,~\\
\eta_2&=\eta_{L_1}+\tilde{\theta}_2+(s-L_1)\eta'_{L_1}
,~
\eta'_2=\theta_2+\eta'_{L_1}\;\;,
\end{aligned}
\end{equation}
where the bending angles and their integrals, using Eq.~\eqref{eq:th}, are expressed as:
\begin{equation}
\theta_{1}=\mathlarger{\int\limits}_0^{s} \dfrac{1}{\rho_{1}(s)} ds,\hspace{0.1cm} \theta_2=\mathlarger{\int\limits}_{L_{1}}^{s} \dfrac{1}{\rho_{2}(s)} ds, \hspace{0.1cm} \tilde{\theta}_{1}=\mathlarger{\int\limits}_0^{s} \theta_{1} ds, \hspace{0.1cm} \tilde{\theta}_{2}= \mathlarger{\int\limits}_{L_{1}}^{s} \theta_{2} ds\;\;, 
\end{equation}
The bending angle of the half dipole is then given by: 
\begin{equation}
\label{eq:halftheta}
\theta=\theta_{{1}(s=L_1)}+\theta_{{2}(s=L_1+L_2)}=\mathlarger{\int\limits}_0^{L_1} \dfrac{1}{\rho_{1}(s)} ds+ \mathlarger{\int\limits}_{L_{1}}^{L_1+L_2} \dfrac{1}{\rho_{2}(s)} ds\;\;.
\end{equation}
Inserting the partial dispersion invariants into Eq.~\eqref{eq:epsi}, the emittance is found to be:
\begin{equation}
\label{eq:ex}
\epsilon_x=G \left(\dfrac{1}{L_1} \mathlarger {\int\limits}_0^{L_{1}} \dfrac{{\cal H}_1}{\left| \rho_1(s)  \right|^3 }ds+\dfrac{1}{L_2} \mathlarger {\int\limits}_{L_{1}}^{L_{1}+L_{2}} \dfrac{{\cal H}_2}{\left| \rho_2(s)  \right|^3 }ds \right), ~ \mathrm{where} ~~G=\dfrac{C_{q} \gamma^2}{J_{x}}\left(\dfrac{1}{L_1}  \mathlarger{\int\limits}_0^{L_{1}} \dfrac{1}{\rho_{1}(s)^2}ds + \dfrac{1}{L_2} \mathlarger {\int\limits}_{L_{1}}^{L_{1}+L_{2}} \dfrac{1}{\rho_{2}(s)^2} ds \right) ^{-1}.
\end{equation}
The final form of Eq.~\eqref{eq:ex} 
can be expressed as:
\begin{equation}
\label{eq:exelegant}
\epsilon_x=G \dfrac{I_7+I_8\lambda+(I_1+I_2\lambda)\beta_{cd}^2+\eta_{cd}\left(I_5+I_6\lambda+(I_3+I_4\lambda)\eta_{cd}\right)}{L_1 \beta_{cd}}\;\;,
\end{equation} 
with the integrals $I_1-I_8$ given in Appendix~\ref{integrals}.

In order to find the absolute minimum emittance, the first partial derivatives of the emittance with respect to the beta and dispersion functions should be zeroed. As a result, the $\beta_{TME}$ and $\eta_{TME}$ values that achieve the TME at the center of the dipole are found to be:
\begin{equation}
\label{eq:betaeta}
\beta_{TME}= \dfrac{\sqrt{-(I_5+I_6\lambda)^2+4(I_7+I_8\lambda)(I_3+I_4 \lambda)}}{2\sqrt{(I_1+I_2\lambda)(I_3+I_4\lambda)}} \hspace{0.2cm} \mathrm{and} \hspace{0.2cm}
\eta_{TME}= -\dfrac{I_5+I_6\lambda}{2(I_3+I_4\lambda)}\;\;.
\end{equation}
By inserting Eq.~\eqref{eq:betaeta} into Eq.~\eqref{eq:exelegant} the expression for the TME is given by:
\begin{equation}
\label{eq:etme}
\epsilon_{TME}=G \dfrac{\sqrt{I_1+I_2\lambda}\sqrt{-(I_5+I_6\lambda)^2+4(I_7+I_8\lambda)(I_3+I_4 \lambda)}}{L_1\sqrt{I_3+I_4\lambda}}~.
\end{equation}
The emittance detuning factor $\epsilon_r$ that describes the deviation of the emittance $\epsilon_{x}$ from its theoretical minimum $\epsilon_{TME}$ is given by: 
\begin{equation}
\label{eq:er}
\epsilon_r=\dfrac{\epsilon_{x}}{\epsilon_{TME}}= \dfrac{\sqrt{I_3+I_4\lambda}\left(I_7+I_8\lambda+(I_1+I_2\lambda)\beta_{cd}^2+\left(I_3 + I_4\lambda\right) \eta_{cd}^2-\left(I_5 + I_6\lambda\right)\eta_{cd}\right)}{\beta_{cd}\sqrt{I_1+I_2\lambda}\sqrt{-(I_5+I_6\lambda)^2+4(I_7+I_8\lambda)(I_3+I_4 \lambda)}}
\end{equation}
Solving Eq.~\eqref{eq:er} with respect to $\beta_{cd}$ gives:
\begin{equation}
\label{eq:betacd}
\beta_{cd_{1,2}} = 
\dfrac{ \epsilon_r\sqrt{4  \left(I_7 + I_8\lambda\right)-\dfrac{\left(I_5 + I_6\lambda\right)^2}{I_3 + I_4\lambda }}\pm  \sqrt{4 \left( \left(I_7 + I_8\lambda\right) \left(\epsilon_r^2-1\right)+ \left(I_3 + I_4\lambda\right) \eta_{cd}^2-\left(I_5 + I_6\lambda\right)\eta_{cd} \right) -\dfrac{\left(I_5 + I_6\lambda\right)^2}{I_3 + I_4\lambda} \epsilon_r^2} }{2\sqrt{I_1+I_2\lambda}}
\;\;.
\end{equation}
Applying the requirement of $\beta_{cd}$ to be a real-positive number, the quadratic dependence of the argument in the square root on the dispersion at the center of the dipole must have an upper and a lower limit, i.e.  $\eta_{min}<\eta_{cd}<\eta_{max}$, given by: 
\begin{equation}
\label{eq:etacd}
\eta_{cd_{min,max}}=-\dfrac {(I_5 + I_6\lambda) \pm \sqrt{((I_5 + I_6\lambda)^2 - 4 (I_3 + I_4\lambda) (I_7 + I_8\lambda)) (1 -\epsilon_r^2)}}{2(I_3+I_4\lambda)}\;\;.
\end{equation}
The $\beta_{cd}$ has two solutions for a fixed $\eta_{cd}$. The solutions of $\beta_{cd}$ and $\eta_{cd}$,  that depend on the detuning factor $\epsilon_r$, determine the limits of $\epsilon_x$.  


\par The horizontal and vertical phase advances of the cell can be found using the trace of the cell transfer matrix. The horizontal phase advance, using Eq.~\eqref{eq:g}, is given by:
\begin{eqnarray}
\label{eq:cosphix}
\cos \mu_x = & \dfrac{g_2^2 - \beta_{cd}^2 \theta^2}{g_2^2 + \beta_{cd}^2 \theta^2}\;\;.
\end{eqnarray}
For a uniform dipole, the horizontal phase advance in order to reach the absolute minimum emittance is independent of any cell or dipole characteristics and has the unique value $\mu_x=284.5^{\circ}$~\cite{ref:streun}. However, in the case of the non-uniform dipoles,  the horizontal phase advance for reaching the TME condition depends on $\theta$ and $\tilde{\theta}$ and thereby, on $\tilde{\rho}$ and $\lambda$. 
The vertical phase advance can be expressed as:
 \begin{equation}
 \label{eq:cosphiy}
 \begin{split} 
 \cos \mu_y = & 1+(L+2(s_{13}+s_{12}))(\frac{1}{f_1}+\frac{1}{f_2})+2\frac{(L+s_1+s_{12})s_{23}+s_{12}s_3}{f_1 f_2}
+\frac{(L+2s_1)s_{23}}{f_1^2} \\  & +\frac{(L+2(s_1+s_{12}))s_3}{f_2^2}+\frac{(L+2s_1)(s_2^2+2s_2s_3)}{f_1^2f_2}
 +2\frac{(L+2s_1+s_{12})s_2s_3}{f_1f_2^2}+\frac{(L+2s_1)s_2^2s_3}{f_1^2f_2^2}\;\;,
 \end{split}
 \end{equation}
where $s_{12}=s_1+s_2$, $s_{13}=s_1+s_3$ and $s_{23}=s_2+s_3$. If the cell is tuned to the absolute minimum emittance conditions, for $s_1 \to 0$ or $s_2 \to 0$ or $s_3 \to 0$ and based on the results presented in Appendix~\ref{focal}, the $\cos \mu_y$ goes to infinity and so, the vertical motion is unstable. Unlike the horizontal phase advance, the vertical one depends both on the optics functions at the dipole center and on the cell geometry.

\section{DIPOLE PROFILES}
\label{dipprof}
\par
\begin{figure}[h]
\centering
\includegraphics[width=0.35\linewidth]{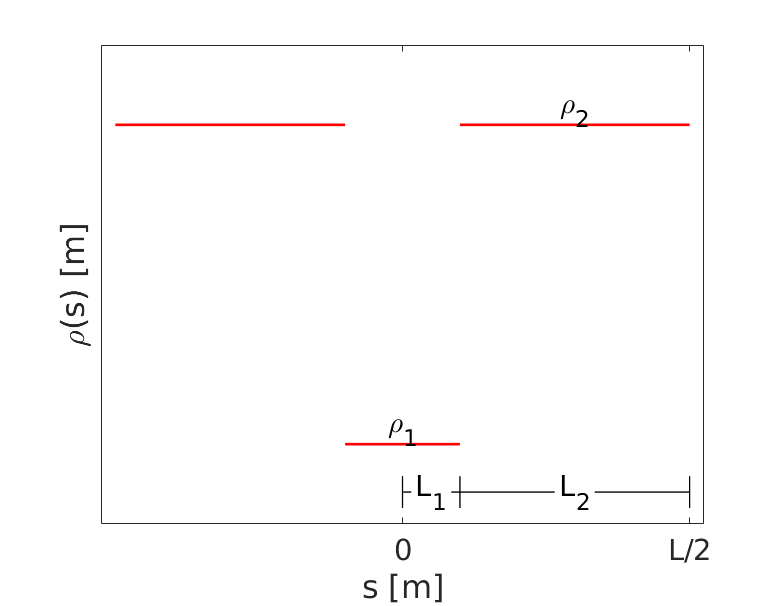}
\quad
\includegraphics[width=0.35\linewidth]{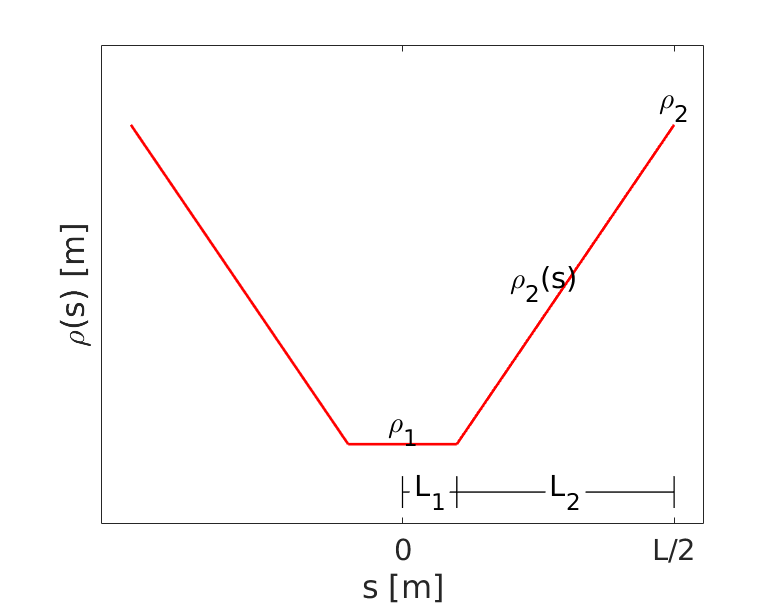}
\caption{\small {The evolution of the beding radius along the step (left) and the trapezium (right) dipole profile.}}
\label{fig:step}
\end{figure}

 Based on studies of preceding works for the longitudinally variable bends~\cite{ref:gr,ref:nw,ref:pe,ref:esrf,ref:wwp,ref:wang,ref:ler14_stef, ref:streun15,ref:ipac15_stef}, two dipole profiles are presented where the bending radius forms a step and a trapezoidal shape. 
The step profile shown in Fig.~\ref{fig:step} (left) consists of two different constant field segments, having the minimum bending radius at the dipole center. The evolution of $\rho$ for the step profile is given by:
\\
$\rho(s)= 
\begin{array}{cc}
  \Bigg\{ & 
    \begin{array}{l}
     \rho_{1},~ 0<s<L_{1}  \\
     \rho_{2},~ L_{1}<s<L_{1}+L_{2}=L/2
    \end{array}
\end{array}
$
\\
The trapezium profile is shown in Figure~\ref{fig:step} (right), where again the strongest
constant field  segment is localized at the center of the dipole.
The evolution of the bending radius from the dipole center until its edge is expressed as:
\\
$\rho(s)=$ $
\begin{array}{cc}
  \Bigg\{ & 
    \begin{array}{l}
      \rho_{1} ,~ 0<s<L_{1}  \\
     \rho_{1}+(L_1-s)(\rho_{1}-\rho_{2})/L_2 ,~ L_{1}<s<L_{1}+L_{2}=L/2
    \end{array}
\end{array}
$\\
\par  The theoretical minimum emittance, as calculated using Eq.~\eqref{eq:etme} for each dipole profile, depends on $\tilde{\rho}$, $\lambda$ and $\theta$. 
The  reduction factor that describes the reduction of the minimum emittance for a non-uniform dipole with respect to a uniform one with the same bending angle, is defined as: 
\begin{equation}
\label{eq:reduct}
F_{TME}=\dfrac{\epsilon_{TME_{uni}}}{\epsilon_{TME_{var}}}\;\;,
\end{equation} 
where $\epsilon_{TME_{uni}}$ and $\epsilon_{TME_{var}}$ are the TMEs for a uniform dipole and a variable bend respectively. The $F_{TME}$ depends only on $\tilde{\rho}$ and $\lambda$. For both dipole cases, the full expression of the $F_{TME}$ is given in Appendix~\ref{redfactor}. 
In practice, the TME cells are detuned to reach larger emittances so that the cell characteristics are more relaxed. Moving away from the TME, the resulted emittances are $\epsilon_{var}$ and $\epsilon_{uni}$ for the non-uniform and for the iso-magnetic dipoles, respectively. In order to compare the emittances of a uniform and of a non-uniform bending magnet, their ratio (using Eqs.~\eqref{eq:er} and \eqref{eq:reduct}) is defined as:
\begin{equation}
\label{eq:erftme}
\dfrac{\epsilon_{var}}{\epsilon_{uni}}=\dfrac{\epsilon_{r_{var}}\epsilon_{TME_{var}}} {\epsilon_{r_{uni}}\epsilon_{TME_{uni}}}= \dfrac{\epsilon_{r_{var}}}{\epsilon_{r_{uni}}} \dfrac{1}{F_{TME}}\;\;,
\end{equation}
where $\epsilon_{r_{var}}$ and $\epsilon_{r_{uni}}$ are the detuning factors for the non-uniform and for the uniform dipole respectively. In order to get an emittance reduction, it should always be  $\dfrac{\epsilon_{var}}{\epsilon_{uni}}<1$. Thus, using Eq.~\eqref{eq:erftme}, the restriction of:
\begin{equation}
\label{eq:restr}
 \dfrac{\epsilon_{r_{var}}}{\epsilon_{r_{uni}}}<F_{TME}
\end{equation}
is established. The smaller is the ratio of the detuning factors compared to the $F_{TME}$ (that is fixed in accordance to the chosen dipole characteristics), the lower is the final emittance the variable bend gives. Practically this means that even if the detuning of a TME cell with a variable bend is larger than in the case of using a uniform dipole, emittance reductions are possible if Eq.~\eqref{eq:restr} is satisfied. 

\par The characteristics of a realistic dipole profile are driven by the constraint of how sharply and quickly the transition from the high to the low field can be established. Regarding the fact that the fringe field of the first dipole part should not significantly affect the field of the second one and that a sharp field drop off is technologically questionable, the minimum difference between the highest and the lowest field and the corresponding  difference in the lengths is assumed to be $\lambda, \tilde{\rho}>0.04$. 
Based on the optimal variable bend characteristics, a design study is on-going in order to provide the final specifications of the dipole to be fabricated~\cite{ref:MagnetDesign1,ref:MagnetDesign2}.
\begin{figure}[h]
	\centering
	\includegraphics[width=0.33\linewidth]{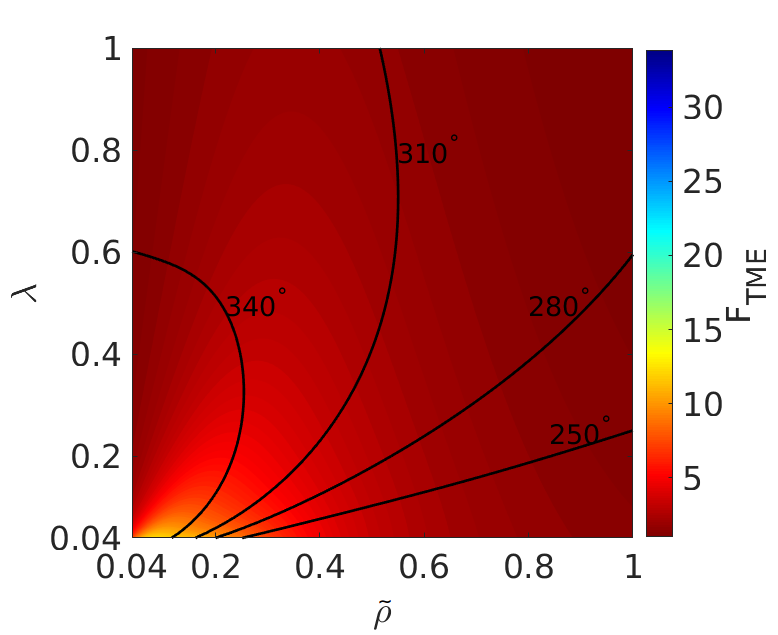}
	\quad
	\includegraphics[width=0.33\linewidth]{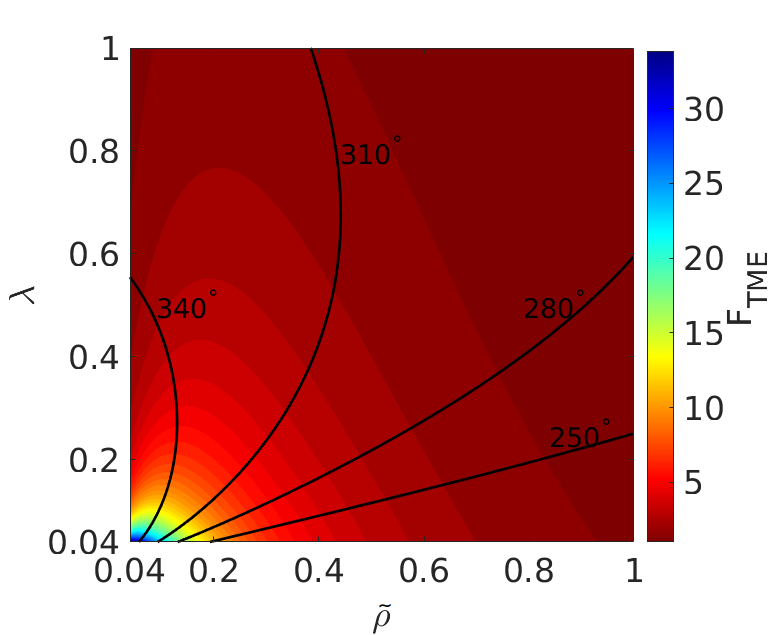}
	\captionsetup{justification=raggedright,singlelinecheck=false}
	\caption{\small {The parametrization of the reduction factor $F_{TME}$ with the ratio of the bending radii $\tilde{\rho}$ and lengths $\lambda$  for the step (left) and the trapezoidal (right) profile, for the TME case. The black contour lines correspond to different values of horizontal phase advances.}}
	\label{fig:wig}
\end{figure}
\par In Figure~\ref{fig:wig}, the reduction factor $F_{TME}$ is parametrized with $\tilde{\rho}$ and $\lambda$, for the step  (left)  and  the trapezoidal (right) profile.  The areas where  $F_{TME}$ is high are blue-colored, while red-colored are the areas where the reduction is smaller. 
The black contour lines show different values of the horizontal phase advance.
As mentioned above, for a uniform dipole, in order to reach the absolute minimum emittance, there is a unique horizontal phase advance of $\mu_x=284.5^{\circ}$~\cite{ref:streun}, independent of any cell or dipole characteristics. However, in the case of the non-uniform dipoles,  the horizontal phase advance for reaching the TME condition depends on $\tilde{\rho}$ and $\lambda$. The highest reductions correspond to high phase advances $\mu_x>310^{\circ}$. Remarkable emittance reductions are reached even for lower phase advances which correspond to smaller chromaticities. 
\par  For both profiles, in the limits where $\tilde{\rho}$, $\lambda\rightarrow1$ (i.e. $\rho_2=\rho_1$ and $\lambda_2=\lambda_1$) there is no emittance reduction. 
In the limits where $\lambda\rightarrow0$ and $\tilde{\rho}\rightarrow0$ (i.e. $L_2>>L_1$ and $\rho_2>>\rho_1$) the reductions obtained are practically infinite.
The highest possible reductions are found to be around 13 and 34 for the step and the trapezium profile, respectively, for $\lambda, \tilde{\rho}>0.04$. These reductions are localized where both $\lambda$ and $\tilde{\rho}$ are low, demanding a sharp transition from the high to the low field region.   
The issue of concern for the design of a variable bend is how small can $\tilde{\rho}$ be in order to get a realistic difference between the maximum and the minimum magnetic field along a specific dipole length that has a fixed bending angle. 
\par In order to facilitate the comparison between the step and the trapezium profile, the number of dipoles $N_d$, their length $L$ and the minimum bending radius $\rho_1$ values are kept the same. As a numerical example, the minimum $\rho_1$ value is set to $4.1$~m, i.e. $B=2.3$~T at an energy of $2.86$~GeV for the CLIC DRs. In addition, examples for dipole lengths and angles different than the ones of the CLIC DR design are presented. Using Eq.~\eqref{eq:halftheta}, the bending angles for the step and the trapezium profile respectively, are found to be:
\begin{equation}
\label{eq:angles}
\theta_{step}=\dfrac{L (\lambda + \tilde{\rho})}{\rho_1(1 + \lambda)}
~~~\mathrm{and}~~~ 
\theta_{trapezium}=\dfrac{L (\lambda (-1 + \tilde{\rho}) + \tilde{\rho} \ln\tilde{\rho})}{\rho_1 (-1 + \tilde{\rho}) (1 + \lambda)}\;\;.
\end{equation} 
Solving Eq.~\eqref{eq:angles} with respect to the ratio $\tilde{\rho}$, for a fixed bending radius $\rho_1$, provides an expression of $\tilde{\rho}$ depending on the bending angle $\theta$, the dipole length $L$ and the lengths ratio $\lambda$. Thereby, the reduction factor of Eq.~\eqref{eq:reduct} becomes a function of the bending angle $\theta$ (or the number of bends $N_d$), the dipole length $L$ and the lengths ratio $\lambda$. Even if the reduction factors achieved get higher by increasing the number of dipoles and their length, a compromise between these two parameters is required. Additionally, as shown in the next section, the fabrication of a variable bend sets a lower limit on the length ratio $\lambda$~\cite{ref:MagnetDesign2} and thus, an upper limit on the reduction factor values that can be achieved.
\begin{figure}[h]
	\centering
	\includegraphics[width=0.33\textwidth]{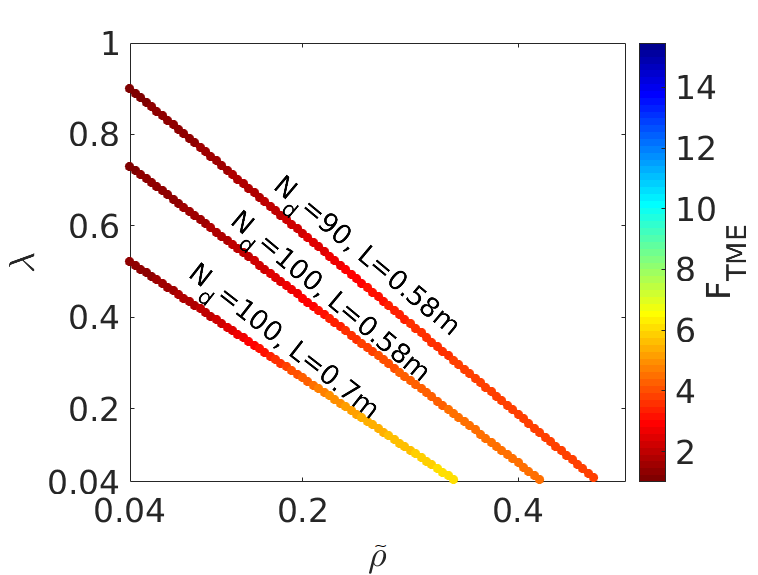}
	\quad
	\includegraphics[width=0.317\textwidth]{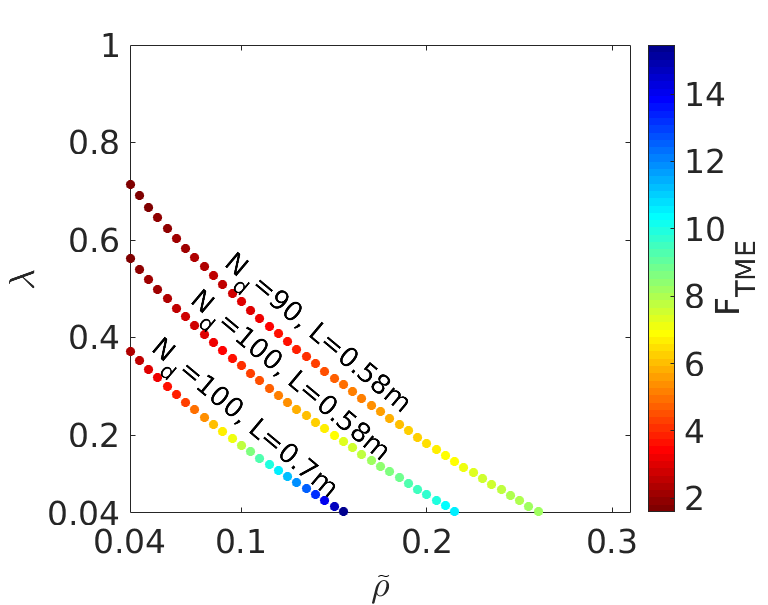}
	\captionsetup{justification=raggedright,singlelinecheck=false}
	\caption{\small {The parametrization of the $F_{TME}$ with $\tilde{\rho}$ and $\lambda$, when fixing $N_d$ and  $L$, for  the step (left) and  the trapezium (right) profile.}}
	\label{fig:ftmescatter}
\end{figure}
\par
\begin{figure}[h]
	\centering
	\includegraphics[width=0.38\textwidth]{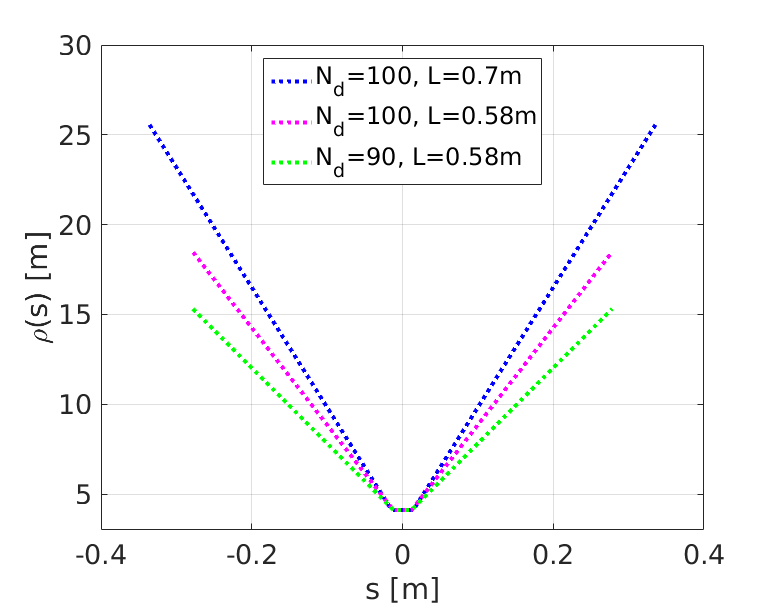}
	\captionsetup{justification=raggedright,singlelinecheck=false}
	\caption{\small {The variation of the bending radius along the dipole length for different ($N_d$, $L$) pairs, resulting in the highest emittance reduction (where $\lambda = 0.04$) for each case, for the case of the trapezium profile.}}
	\label{fig:trapshapes}
\end{figure}

 After imposing the bending angle $\theta$, the dipole length $L$ and minimum bending radius $\rho_1$ in Eq.~\eqref{eq:angles}, a relationship between the ratios $\tilde{\rho}$ and $\lambda$ is obtained and can be inserted in Eq.~\eqref{eq:reduct} for obtaining the respective emittance reduction factors. In this way, the maximum emittance reductions for fixed dipole characteristics can be found. 
Figure~\ref{fig:ftmescatter} shows the parameterization of the reduction factor $F_{TME}$ with the ratios $\tilde{\rho}$ and $\lambda$, for both dipole profiles and again with the restriction of $\lambda,\tilde{\rho}>0.04$, for three different cases of dipole numbers and lengths ($N_d$, $L$) equal to ($100$, $0.7$~m), ($100$, $0.58$~m) and ($90$, $0.58$~m). The case corresponding to $100$ dipoles with a length of $0.58$~m matches the exact dipole characteristics of the CLIC DRs. For the dipole constraints applied in each case, the trapezium gives always higher reductions than the step profile. Obviously, the more and the longer the dipoles are, the higher emittance reductions are achieved.  The maximum reductions in all cases are localized where both $\lambda$ and $\tilde{\rho}$ are low. Large length ratios are not of interest since the reduction factor becomes very small. The bending radius ratios resulting in the highest emittance reductions are lower for the trapezoidal profile. The variation of the bending radius along the dipole length that results in the highest emittance reduction for each case, is plotted in Fig.~\ref{fig:trapshapes} for the trapezium profile. As was expected from Fig.~\ref{fig:ftmescatter}, the maximum emittance reductions (where $\lambda = 0.04$) correspond to different ratios $\tilde{\rho}$ depending on the number of dipoles and length and thus, for a fixed minimum bending radius ($\rho_1$=$4.1$~m), their maximum bending radius differs.

\section{Longitudinally variable field dipole design for the CLIC Damping Rings}
\label{magdesign}
\par  The analytical results showed that the trapezium dipole profile can reach very low emittances, compared to a uniform dipole of the same bending angle.  Therefore, the fabrication of a variable bend having a bending radius that forms a trapezium shape is of interest. According to the optimal characteristics  of a trapezium dipole profile to be used for the CLIC Damping Rings, the magnetic design of a longitudinally variable bending dipole based on permanent magnets was studied and the prototype will be fabricated by CIEMAT~\cite{ref:MagnetDesign1, ref:MagnetDesign2}.  
\begin{figure}[h!]
		\includegraphics[width=0.43\textwidth]{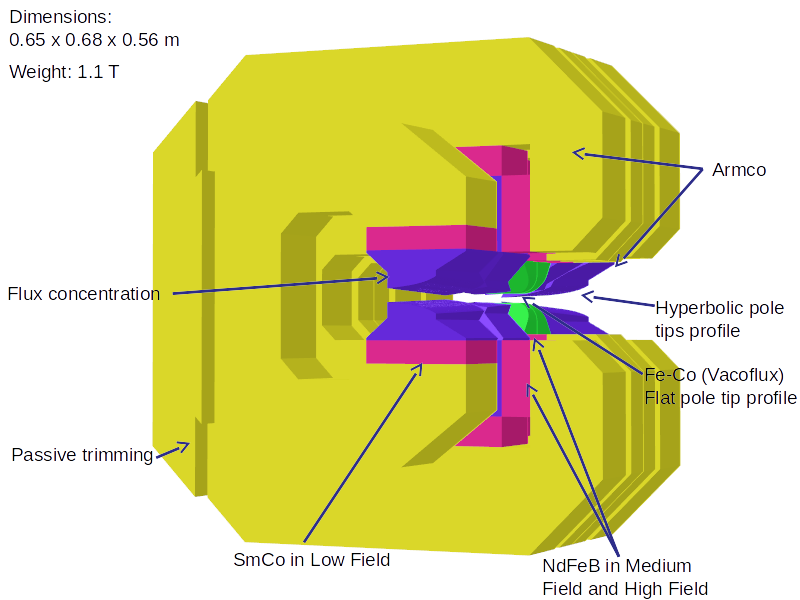}
		\captionsetup{justification=raggedright,singlelinecheck=false}
		\caption{{\small Magnet design based on the characteristics of the variable bends for the CLIC DRs~\cite{ref:MagnetDesign1, ref:MagnetDesign2}.}}
	\label{fig:magnet_design1}
\end{figure}
The main challenges of this design are the bending radius variation which should change linearly along the magnet and the high field region length that is very short. The longitudinal gradient with a trapezoid decay is solved by splitting the magnet in three differentiated field regions combined with an innovative variable gap solution, as presented in Fig.~\ref{fig:magnet_design1}.
\begin{figure}[h!]
		\includegraphics[width=0.36\textwidth]{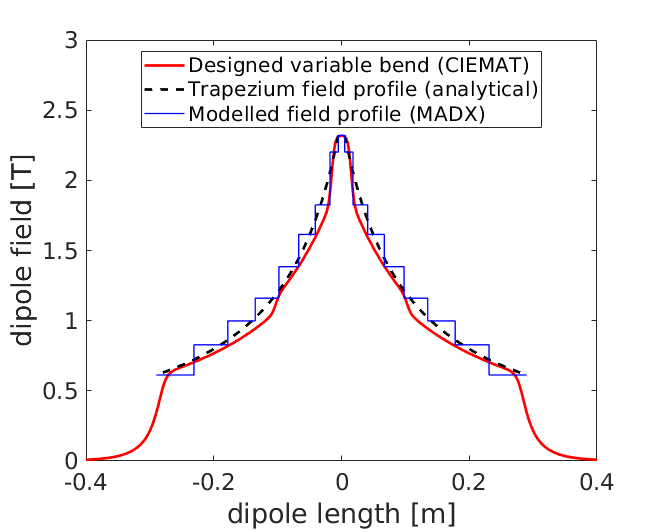}
		\captionsetup{justification=raggedright,singlelinecheck=false}
		\caption{{\small The field variation along a dipole having a peak field of 2.3~T; the designed trapezium profile (red colored), the resulted from analytical calculations profile (black colored) and the simulated in MADX field profile (blue colored).}}
	\label{fig:magnet_design2}
\end{figure}
The low field block is made of SmCo magnets. The medium field has the same configuration as the high field section, using NdFeB blocks. The requested peak field was initially limited to 1.77~T as a reasonable value for a non-superconducting magnet, requiring to deal with iron saturation that is partially overcome using a Fe-Co material and suppressing the hyperbolic profile in the high field region pole tip which is the most saturated section.  
\par The higher is the magnetic field at the center of a variable bend, the higher emittance reduction factors can be reached. The use of permanent magnets could allow having a higher field than in the case of a normal conducting magnet.  The 3D simulations performed have shown that the peak field could be increased above 2~T, resulting in higher emittance reduction factors.  In Fig.~\ref{fig:magnet_design2}, a 2.3~T designed trapezium profile is shown in red. The field decay successfully matches the desired from the hyperbolic field profile provided by the analytical estimates (black colored).  For the designed trapezium profile, the $\lambda$ and $\tilde{\rho}$ values reached are respectively 0.036 and 0.295, corresponding to an emittance reduction factor of $F_{TME}=7.1$~\footnote[2]{The technological restrictions do not allow $\tilde{\rho}$ to reach the optimal value of 0.263 for a $\lambda=0.036$, providing a reduction factor $F_{TME}=8.3$.}.   For the beam optics simulations performed with MADX~\cite{ref:madx}, the field of the designed trapezium profile is being approximated by a sequence of dipoles with a step-like field (blue colored). Finally, since a transverse gradient of -11 T/m was requested, the designed magnet provides at the same time dipolar and quadrupolar fields, having pole tips with a hyperbolic profile to produce the gradient.

\section{Numerical application for  a VARIABLE BEND TME CELL}
\label{vbtme}
In this section, a numerical application of the aforementioned analytical results are presented for the CLIC DR TME cell. The DRs provide the final stage of damping of the $e^+e^{{-}}$ beams of the linear collider after the pre-damping rings and their lattice has a racetrack shape with two arc sections and two long straight sections (LSS), with the TME cells being placed in the arcs~\cite{ref:clic}. The required output horizontal normalized emittance for the CLIC DR, at the energy of $2.86$~GeV, is $500$~nm-rad, i.e. around 90~pm geometrical emittance. According to the design of a variable bend, the maximum dipole field is set to be $2.3$~T (corresponding to the minimum bending radius $\rho_{1}=4.1$~m), for dipoles having a length of $L=0.58$~m. The maximum pole tip field of the quadrupoles and the sextupoles is $B_q^{\mathrm{max}}=1.1$~T and $B_s^{\mathrm{max}}=0.8$~T respectively. 
By fixing these quantities, the free parameters left are the drift space lengths $s_1$, $s_2$, $s_3$ and the emittance, based on optics stability and feasibility constraints described below. It should be stressed that,
it is always preferable to carefully detune the cell from the TME to larger emittances (see Eq.~\eqref{eq:restr}), so that the optics
characteristics of the cell and in particular chromaticity is under control.

\subsection{Stability and feasibility constraints}
\label{subscetion:stabfeas}
The optics stability criterion, for both horizontal and vertical planes is given by:
\begin{equation}
\label{eq:phiyy}
|cos\mu _{x,y}|  <1\;\;.
\end{equation}
The pole tip field value of both quadrupoles and sextupoles has a limit and the radius of the magnet aperture has a minimum value. The feasibility of the quadrupoles is ensured if the quadrupole strength $k$ is kept below a maximum value given by:
\begin{eqnarray}
k=\dfrac{1}{f l_q}= \leq \dfrac{1}{(B\rho)} \dfrac{B_q^{\mathrm{max}}}{R_{\mathrm{min}}}\;\;,
\label{eq:gcon}
\end{eqnarray}
where $B\rho$ is the magnetic rigidity and $B_q^{\mathrm{max}}$ is the quadrupole pole tip field. The  $
R_{\mathrm{min}} = \sqrt{\beta \epsilon_{\mathrm{max}} + ((\frac{\delta p}{p_0})_{\mathrm{max}}D)^2}$  is the minimum required aperture radius for a Gaussian beam distribution, where $\epsilon_{\mathrm{max}}$ is the maximum value of the Courant-Snyder invariant for the injected beam and $(\delta p/p_0)_{\mathrm{max}}$ the corresponding total momentum deviation. As the sextupoles are set to cancel the chromaticity induced by the quadrupoles, their
strength can be calculated by 
$
\xi_{x,y} =  -\frac{1}{4 \pi} \oint{\beta_{x,y}[K_{x,y}(s)-S(s)D(s)]ds}=0
$,
where $K_{x,y}$ correspond to the focusing and defocussing quadrupole strengths. Denoting the pole tip field of the sextupoles as $B_s^{\mathrm{max}}$, their feasibility is ensured if the strength $S$ is lower than a maximum value: 
\begin{eqnarray}
S \leq \dfrac{2 B_s^{\mathrm{max}}}{R_{\mathrm{min}}^2} \dfrac{1}{(B\rho)}\;\;.
\label{eq:Sconst}
\end{eqnarray}

\subsection{Parametrization with the drift lengths}
\begin{figure}[h]
	\centering
	\includegraphics[width=0.33\textwidth]{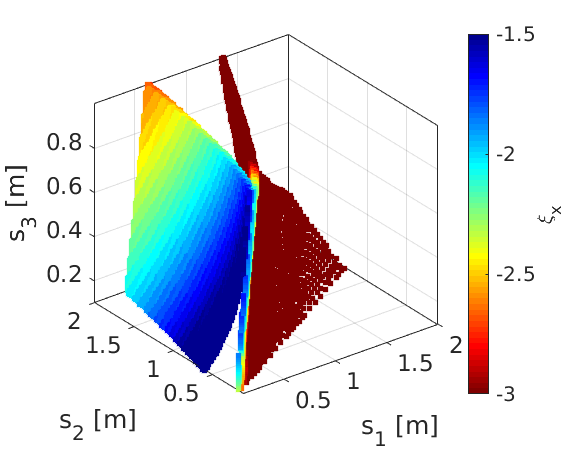}
		\quad
	\includegraphics[width=0.33\textwidth]{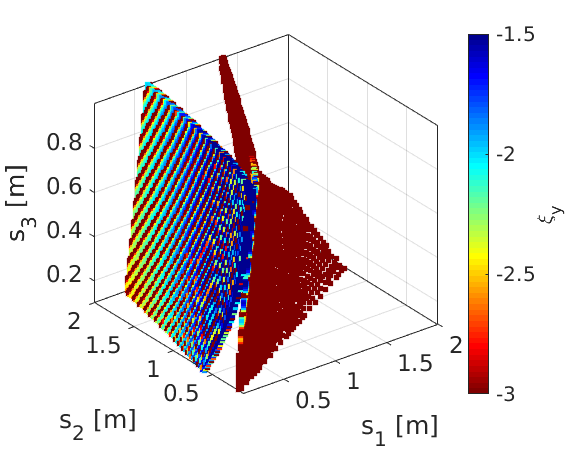}\\
	\includegraphics[width=0.33\textwidth]{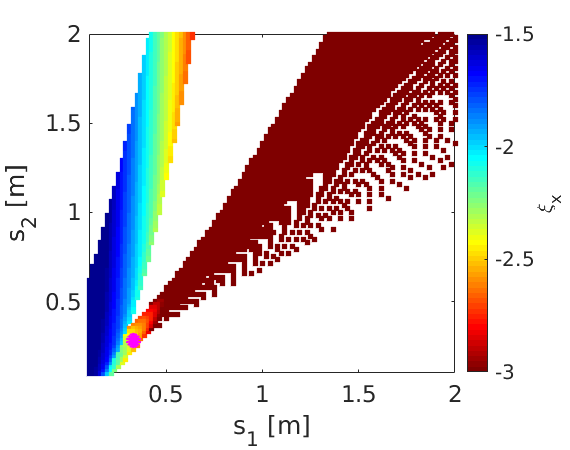}
	\quad
	\includegraphics[width=0.33\textwidth]{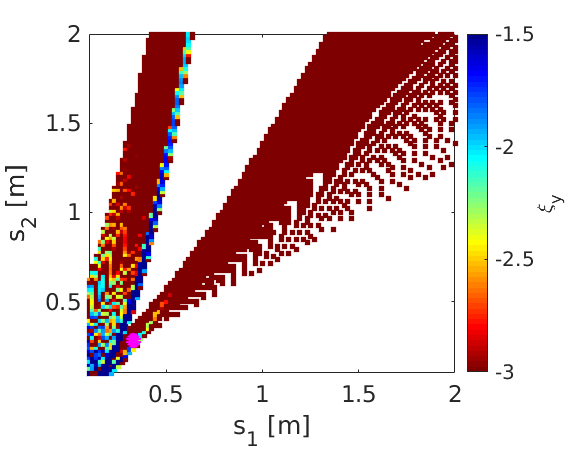}
	\captionsetup{justification=raggedright,singlelinecheck=false}
	\caption{\small {The horizontal (left) and vertical (right) chromaticities as parametrized with the drift lengths $s_1,s_2,s_3$ (top) for the TME with the trapezium dipole profile. In the bottom plot, the same parametrization is plotted and projected to the $s_1,s_2$ plane. High absolute chromaticities correspond to red, whereas low absolute values are blue.}}
	\label{fig:test2}
\end{figure}

The dependence of different cell characteristics on the drift space lengths require their parametrization with $s_1$, $s_2$, $s_3$. A scanning of drift space lengths was performed, aiming to solutions with low chromaticities and small quadrupole strengths,  while keeping the cell compact. 
The chromaticities were calculated for all combinations of drift lengths when  $s_1[m] \in [0.1, 2]$, $s_2[m] \in [0.1, 2]$ and $s_3[m] \in [0.1, 1]$ and the results are presented in Fig.~\ref{fig:test2}. In these plots, the drift lengths are color-coded with the horizontal (left) and vertical (right) chromaticity.
The lowest absolute chromaticities in both planes are found for $s_1<0.5$~m. The absolute minimum emittances exist only for large negative chromaticities per cell, i.e. for $\xi_x<-2$ and $\xi_y<-1$. Detuning factors that give emittances larger than the TME can provide solutions with lower negative chromaticities. Although a careful choice of the first and second drift lengths ($s_1$ and $s_2$) is essential, the cell parameters are less dependent on the third drift length $s_3$. 
In this respect, in order to clearly distinguish the two regions of optically stable solutions, the parametrization of the chromaticities as projected in the $s_1,s_2$ plane is presented in Fig.~\ref{fig:test2} (bottom). The horizontal chromaticity depends strongly on drift length between the two quadrupoles $s_2$. In fact, an area of low horizontal chromaticities can be found, whereas low vertical chromaticities can be obtained following a thin curve. On the other hand, only a small fraction of the ($s_1$, $s_2$) combinations satisfy the requirements of having focal lengths leading to feasible magnets. In this respect, the optimal values indicated with a magenta colored point are found to be $s_1=0.33$ m, $s_2=0.28$ m and $s_3=0.18$~m, slightly on the right side of the lowest chromaticity areas. 

\subsection{Parameterization with the emittance}
\begin{figure}[h]
	\centering
	\includegraphics[width=0.325\textwidth]{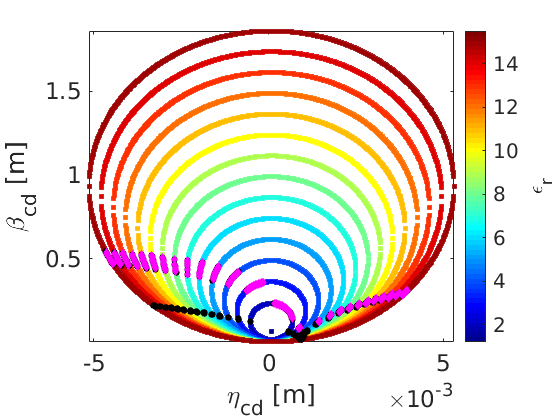}
	\includegraphics[width=0.325\textwidth]{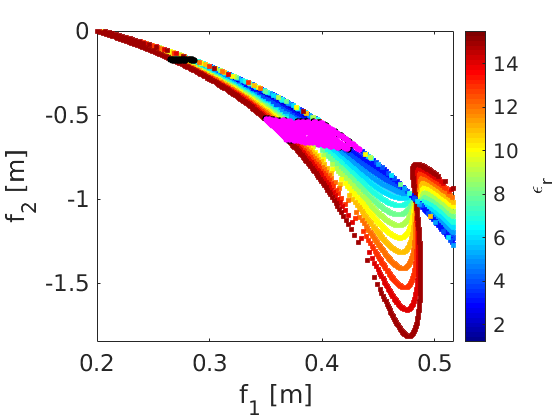}	
	\includegraphics[width=0.325\textwidth]{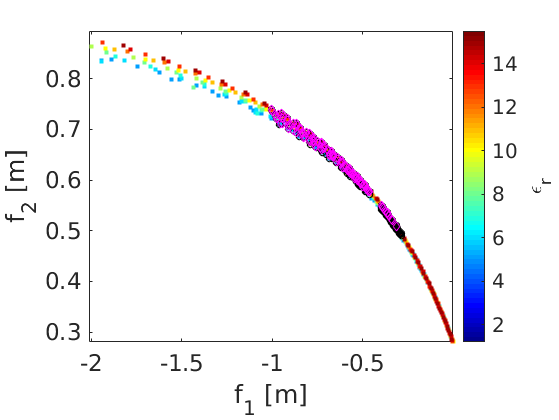}
	\captionsetup{justification=raggedright,singlelinecheck=false}
	\caption{\small {Parameterization of the beta and dispersion functions at the dipole center $\beta_{cd}$, $\eta_{cd}$ (left) and  of the focal lengths, for $f_1>0$, $ f_2<0$ (middle) and for $f_1<0$, $ f_2>0$ (right), with the detuning factor. The black squares indicate stability and the magenta ones feasibility-low chromaticity solutions.}}
	\label{fig:det}
\end{figure}

After fixing the drift lengths, cell properties such as the quadrupole strength can be parametrized with the emittance detuning factor. The emittance value that was so far determined by the reduction factor $F_{TME}$, is increased with this detuning. The emittance reduction due to the non-uniform dipole profile as compared to a standard dipole depends on the relation between the detuning and the reduction factor (see Eq.~\eqref{eq:erftme}). Using this relation, the additional benefit for using longitudinally variable bends instead of the standard ones can be clearly demonstrated. The plots in Fig.~\ref{fig:det} provide the parametrization of the horizontal optics functions at the center of the dipole and the quadrupole strengths color-coded with the detuning factor. Black points correspond to solutions that satisfy the optics stability in both planes. The magenta points are the ones satisfying low chromaticities and quadrupole strengths.
\par The ellipsoidal curves in Fig.~\ref{fig:det} (left) represent the pairs of beta and dispersion functions at the dipole center which result in the same emittance (see Eq.~\eqref{eq:er}). Similarly, the parametrization of the focal lengths with the emittance detuning factor is given in Fig.~\ref{fig:det} (middle) for $f_1>0$ and $ f_2<0$, where the pairs of ($f_1$, $f_2$) lie again on constant emittance curves. Solutions with $f_1<0$ and $f_2>0$ which correspond to the modified TME cell~\cite{ref:modified}, also exist and they are presented in Fig.~\ref{fig:det} (right). 
The TME ($\epsilon_r=1$) is achieved for a unique pair of beta and dispersion functions that is ($\eta_{cd},\beta_{cd})=(1.1\times10^{-4}, 0.07$)~m and  only for one pair of focal lengths which is ($f_1, f_2)=(0.27,-0.18)$~m (out of scale in the bottom right corner of the right part of Fig.~\ref{fig:det}), i.e. quite strong quadrupole strengths are necessary. Furthermore, and for the chosen drift lengths, there are no solutions that assure optics stability and low chromaticities. Solutions that assure lattice stability (black points) and for chromaticities $\xi_x,~ \xi_y>-2.5$ (magenta points) arise when moving away from the TME to larger detuning factors. Solutions with both focal lengths positive are unstable. Even if the chosen cell characteristics result in a confined $\epsilon_r$ region, the low emittances reached for a very compact cell counteracts the fact that it is numerically challenging to tune the cell. One can finally observe that there is a well-determined focal length range in the middle plot of Fig.~\ref{fig:det}, for which the cell can be tuned, satisfying stability and feasibility criteria. For the opposite quadrupole polarity (first quadrupole de-focusing, see right plot of Fig.~\ref{fig:det}), the zone spans a larger focal length range, confined, though, to a narrow line, which may result in a difficulty to reach numerically the desired solution.

\begin{figure}[h]
	\centering
	\subfloat[\label{da}]{\includegraphics[width=0.33\textwidth]{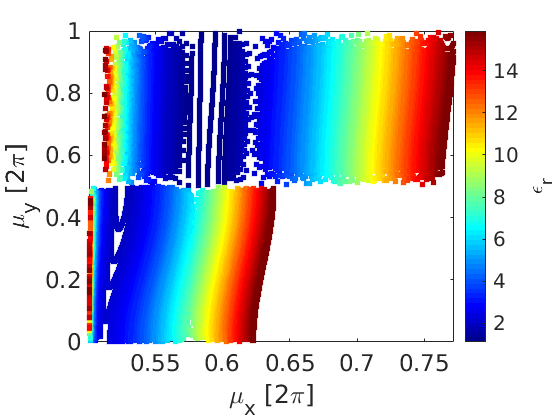}}
	\subfloat[\label{db}]{\includegraphics[width=0.33\textwidth]{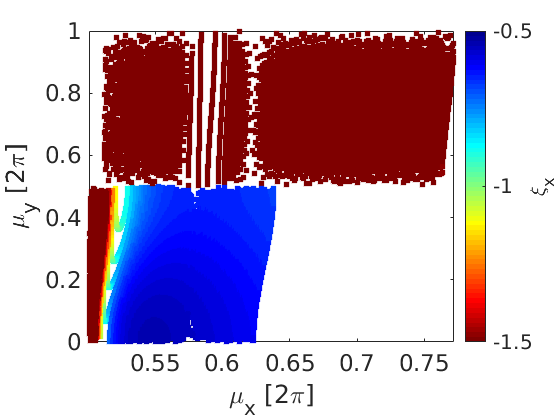}}
	\subfloat[\label{dc}]{\includegraphics[width=0.33\textwidth]{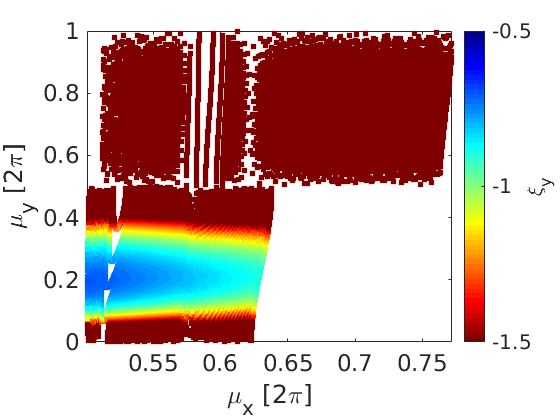}}
	\captionsetup{justification=raggedright,singlelinecheck=false}
	\caption{\small {Parameterization of the detuning factor (a) and the chromaticities $\xi_x$ (b) and  $\xi_y$ (c) with the horizontal $\mu_x$ and vertical $\mu_y$ phase advances, for $f_1>0$, $f_2<0$, for the trapezium dipole profile.}}
	\label{fig:phi}
\end{figure}
\par Figure~\ref{fig:phi} shows the parametrization of the detuning factor $\epsilon_r$ (a) and of the horizontal $\xi _x$ (b)  and vertical  $\xi_y$  (c) chromaticity with the horizontal  $\mu_x$ and vertical $\mu_y$ phase advances, for the case of $f_1>0$, $f_2<0$ solutions which appear only when  $\mu_{x}>0.5\cdot2\pi$.
Towards high vertical phase advances, the chromaticities for both planes have high negative values ($\xi _x$, $\xi_y<-3$). Large horizontal phase advances correspond to minimum dispersion and beta functions at the center of the dipole that require strong focusing and to that end, result in high chromaticities.  It can be noticed that for $\mu_{y}<0.5\cdot2\pi$, there are low negative chromaticities even for small detuning factors corresponding to emittances close to the minimum one. 
\begin{figure}[h]
	\centering
	\subfloat[\label{n}]{\includegraphics[width=0.33\textwidth]{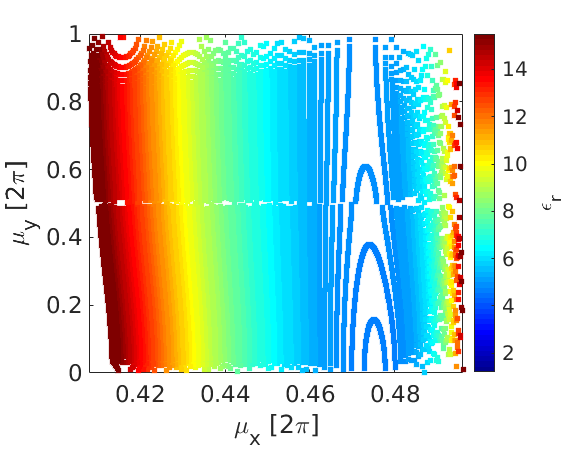}}
	\subfloat[\label{o}]{\includegraphics[width=0.33\textwidth]{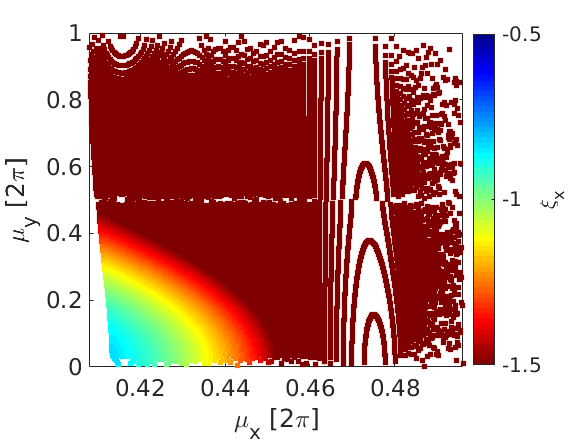}}
	\subfloat[\label{p}]{\includegraphics[width=0.33\textwidth]{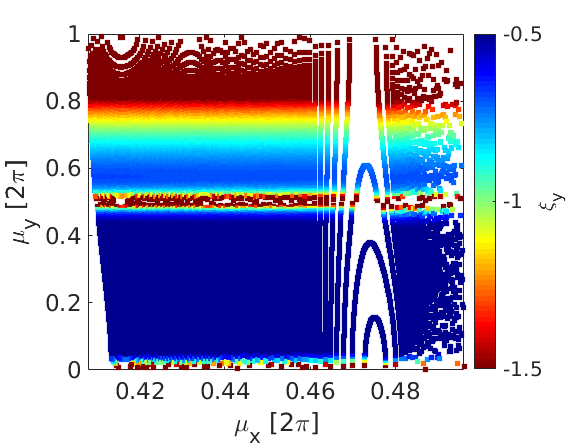}}
	\captionsetup{justification=raggedright,singlelinecheck=false}
	\caption{\small {Parameterization of the detuning factor (a) and the chromaticities $\xi_x$ (b) and  $\xi_y$ (c) with the horizontal $\mu_x$ and vertical $\mu_y$ phase advances, for modified TME cell where $f_1<0$, $f_2>0$, for the trapezium dipole profile.}}
	\label{fig:modified}
\end{figure}
\par Solutions for $f_1<0$, $f_2>0$, corresponding to the modified TME cell~\cite{ref:modified}, also satisfy the stability criteria for the chosen cell characteristics and are presented in Fig.~\ref{fig:modified}. These solutions appear always for  $\mu_{x}<0.5\cdot2\pi$. The quadrupole strengths for the modified TME cell are lower compared to the normal TME cell, as presented in Fig.~\ref{fig:det} (right). Knowing the $F_{TME}$ for the designed variable bend, the restriction described in Eq.~\eqref{eq:restr} sets an upper limit to the detuning of the cell. Since the modified TME cell requires a large detuning in order to get low horizontal chromaticities, the final emittance reductions reached are not sufficient. In this respect, the normal TME cell (having $f_1>0$, $f_2<0$ solutions) will be further used for the following numerical application to the CLIC DR optics design.



\section{Alternative CLIC DR design based on longitudinally variable bends and high field wigglers}
\label{CLICVB}
As discussed in the previous section, the DR lattice~\cite{ref:clic} has a racetrack shape with arc sections composed by TME cells. The two long straight sections (LSS) are composed by FODO cells housing RF cavities, injection and extraction equipment and mainly damping wigglers. 
The DR lattice design is driven by the emittance requirements which for ultra-low values give rise to collective effects~\cite{ref:clic}, with intra-beam scattering (IBS) being the dominant one. An alternative design is proposed, which aims to mitigate the IBS effect for a compact ring, using in the arc cells the designed variable bend presented earlier and an optimized high-field wiggler in the FODO cells. The optimization steps followed, as well as the final parameters for the improved design are discussed in this section.  

\subsection{Optimization of the arc TME cell}
\par For the beam optics simulations performed with the MADX code~\cite{ref:madx}, the field variation of the designed trapezium profile is being approximated by a sequence of dipoles with varying field, as illustrated in Fig.~\ref{fig:magnet_design2}. The dipole length is $L=0.58$ m and the maximum dipole field is $2.3$~T. When the uniform dipoles of the current design are replaced by variable bends, the resulted emittance is lower than the required one. In this case, the subtraction of some TME cells from the existing arc is possible. Actually, the number of dipoles (i.e. total number of TME cells) can be reduced to such an extent that the required emittance is still achieved, thereby resulting in a shorter ring. 
\par As soon as the characteristics of the dipole are fixed, the drift space lengths are chosen in accordance to the results presented in section IV. For a combined function dipole, i.e. having a small defocussing gradient (see section~\ref{magdesign}), the IBS effect is reduced through the increase of the vertical beam size at the center of the bend. Basically, instead of having a low $\beta$ at the center of the dipole in both planes, the optics matching imposes $\beta_y$ to be maximum there. Therefore, there is a reduction of IBS growth rates.  Keeping in mind that a TME cell  with a combined function dipole reduces the IBS effect~\cite{ref:grad}, assists in choosing the proper phase advances, that guarantee low chromaticities and small quadrupole strengths. 
The parametrization with the emittance has shown that the quadrupole strengths for which $f_1>0$ and $f_2<0$ (Fig.~\ref{fig:phi}),  can only be found for $\mu_{x}>0.5\cdot2\pi$.  
A good compromise for the horizontal phase advance is to be around $0.51\cdot2\pi$ and for the vertical phase advances to be always below $0.5\cdot2\pi$. After a detailed scanning of the cell characteristics, the horizontal and vertical phases advances of the TME cell are respectively chosen to be around $0.51\cdot2\pi$ and $0.11\cdot2\pi$. 
Taking into account the emittance reduction and detuning factors, emittances smaller than the ones of the current design with the  uniform dipoles, are reached. 
In this respect, it was possible to reduce the number of dipoles down to 
$N_d=90$ for the case of the designed trapezium profile with 2.3~T maximum field.

\subsection{Optimization of the FODO cell}
\begin{figure}[h]
\centering\includegraphics[width=.33\linewidth]{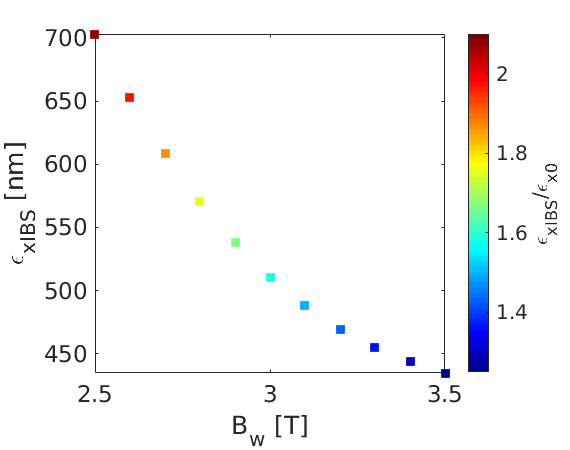}
\quad
\centering\includegraphics[width=.33\linewidth]{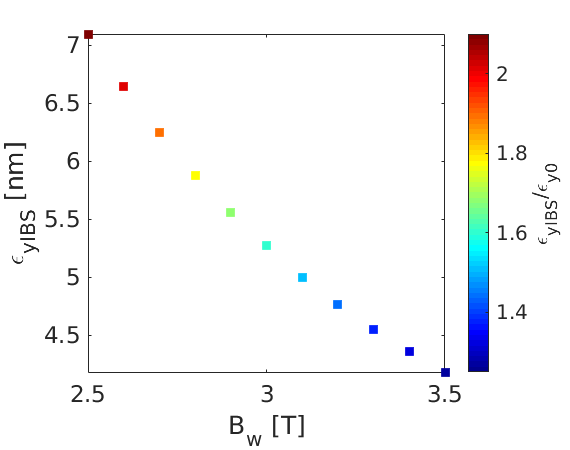}
\captionsetup{justification=raggedright,singlelinecheck=false}
\caption{{\small The dependence of the steady state emittances ($\epsilon_{xIBS}$ and $\epsilon_{yIBS}$) color-coded with their ratio with respect to the corresponding equilibrium emittances ($\epsilon_{x0}$ and $\epsilon_{y0}$) on the wiggler peak field $B_w$, for the  trapezium dipole profile.}}
\label{fig:ibs}
\end{figure}
\par 
Almost every FODO structure of the LSS accommodates two wigglers. The ultimate purpose of using damping wigglers is to further reduce damping times and thereby maintain low emittances by reducing the impact of IBS, but also of various collective effects. At the same time, the fast damping time is imposed by the 50~Hz repetition rate of the collider. 
The use of super-conducting technology is mandatory in order to have a high wiggler field and a relatively short period for obtaining low emittances and fast damping time. The super-conducting magnet wigglers used in the current design have a $B_w=2.5$ T peak field and $\lambda_w=5$ cm period.
\par It was shown that by targeting  higher wiggler fields not only the emittance but also the IBS effect can be reduced~\cite{ref:thesis,ref:wiggler}. Taking into account the optimization of the arc cells and the fact that the emittance with IBS is significantly lowered after increasing the wiggler's peak field, the FODO cells per LSS can be reduced from 13 down to 10. The plots in Fig.~\ref{fig:ibs} show the MADX results of the parametrization of the steady state transverse emittances including the IBS effect with the wiggler peak field $B_w$, starting from the  2.5~T that is the field of the existing wiggler design, for 10 FODO cells. Clearly, the wiggler field increase corresponds to a significant reduction of the IBS effect. The new working point for the damping wigglers, that complies with the technological restrictions, is proposed to be at 3.5~T  and with a 49~mm period length, giving an output emittance of around 435~nm-rad. This design necessitates a different wire technology, using Nb$_3$Sn material~\cite{ref:laura}. 

\subsection{Optical functions and new design parameters}
\begin{figure}[h]
	\centering\includegraphics[width=.378\linewidth]{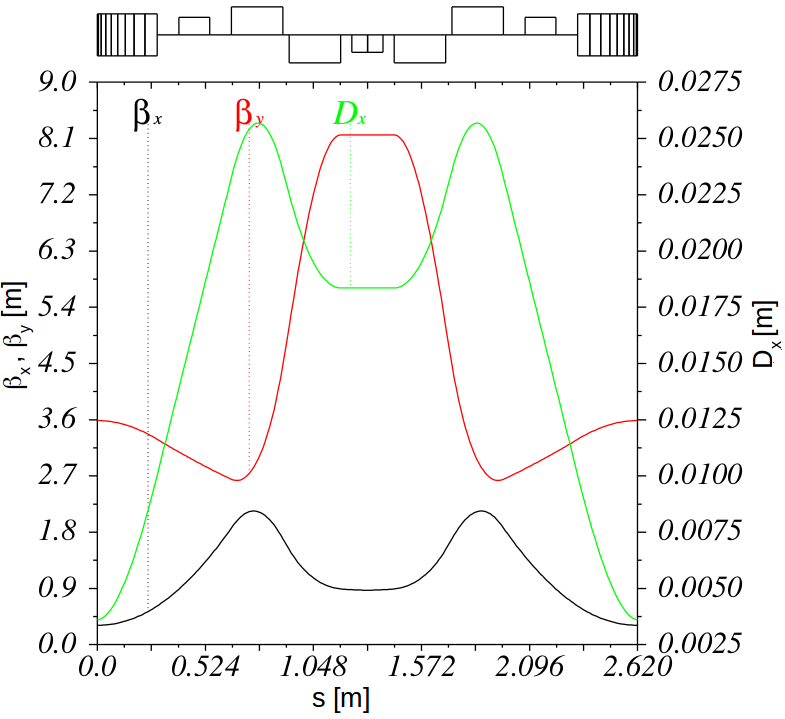}
	\quad
	\centering\includegraphics[width=.4\linewidth]{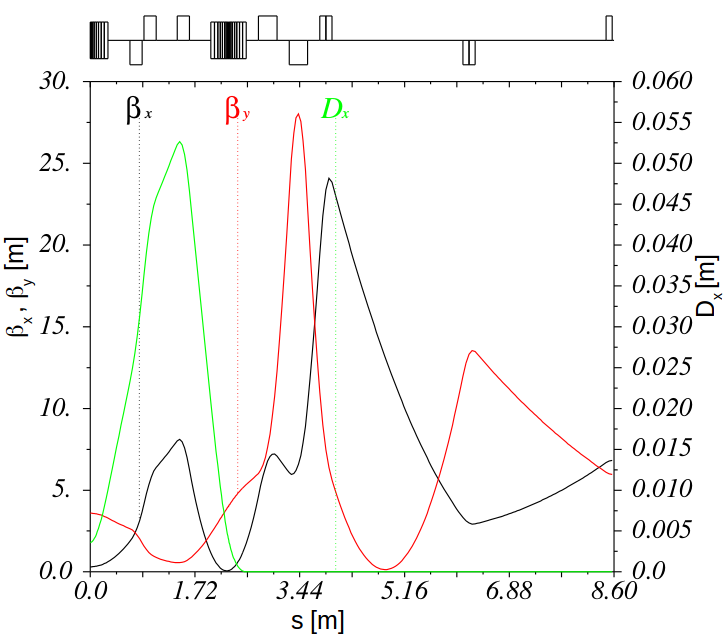}
	\captionsetup{justification=raggedright,singlelinecheck=false}
	\caption{{\small Optical functions of the TME cell (left) and the dispersion suppressor-beta matching section followed by the FODO cell (right), when using in the arcs the trapezium dipole profiles.}}
	\label{fig:tmecells}
\end{figure}
\par  The final lattice, with a smaller number of dipoles and wigglers than the ones of the existing design, is produced. In Fig.~\ref{fig:tmecells} (left), the matched optics, i.e. horizontal dispersion, horizontal and vertical beta functions, are plotted for one arc TME cell. On the top part of the figure, a schematic
layout of the cell is presented, showing the two doublets of quadrupoles and the sextupoles that are placed between the two mirror symmetric defocussing quadrupoles and between the dipole and the focusing quadrupoles. In Fig.~\ref{fig:tmecells} (right), the matched optics of the dispersion suppressor-beta matching section followed by the wiggler FODO cell, are presented.
\par The parameters of the original design and the alternative ones are displayed in Table~\ref{tab:DRparams1}. The alternative designs have 90 dipoles and 40 FODO cells with wigglers of $3.5$~T peak field, while the original has 100 uniform dipoles and 52 FODO cells with wigglers of $2.5$~T peak field. The parameters for the alternative new design are given for the case of uniform and variable bends, keeping the same arc TME cells and FODO cells, so as to have a fair comparison. By having exactly the same arc cells (length, angle and phase advances), the emittance is reduced by $40~\%$ only due to the use of the variable bends. 
For the alternative design, the optimal horizontal phase advance is slightly above 180 deg., resulting in larger quadrupole strengths whose length had to be increased in order to keep the gradient below $100$~T/m and so, the total length of the TME cell became $7~\%$ longer. On the other hand, the chromaticity is significantly smaller for the new DR design. Both the original and the alternative design with variable bends can reach the target emittances including IBS (as calculated by the Bjorken-Mtingwa formalism through MADX~\cite{ref:madx}), with the alternative one giving a $13~\%$ shorter ring circumference. Apart from achieving a lower emittance using the trapezium profile, it is possible to allow significant margin for the $500$~nm target, for an eventual increase of the required bunch population, as lately proposed due to the CLIC re-baselining~\cite{ref:bol}. Indeed, for the case of low energy CLIC which needs a higher bunch charge of $N_b=5.7\times10^9$ the final emittances are increased up to $472$~nm and $4.6$~nm in the horizontal and vertical plane, respectively, still remaining below the design target. 
\begin{table*}
	\begin{center}
		
		\caption{ Design parameters for the original and the improved design of the CLIC DRs, for the case of $N_b=4.1 \times 10^9$}
		\label{tab:DRparams1}
		
		\centering
		{\normalsize \begin{tabular}{lccc}\hline\hline
				
				
				\multirow{2}{*}{\textbf{Parameters, Symbol [Unit]}} 		&\multicolumn{1}{c}{\textbf{Original design}}
				&\multicolumn{2}{c}{\textbf{Alternative design}}\\
				&\textbf{uniform} 	&\textbf{uniform} 	&\textbf{trapezium}  \\	
				\hline  \hline 
				Energy, $E$ [GeV]         		    				&2.86& 	\multicolumn{2}{c}{2.86}  \\     
				Bunch population, $N_b$	[$10^9$]      				&4.07&   \multicolumn{2}{c}{4.07}	  \\                          
				Circumference, $C$ [m]             					& 427.5  & 	\multicolumn{2}{c}{373.7} \\                
				Basic cell type  in the arc/LSS               	    &TME/FODO 	&\multicolumn{2}{c}{TME/FODO} \\
				Number of arc cells/wigglers,	$N_d/N_{w}$			& 100/52 &  \multicolumn{2}{c}{90/40}\\
				RF Voltage,	$V_{RF}$ [MV]				        	& 4.50  & \multicolumn{2}{c}{6.50}\\
				Harmonic number,	$h$        						& 2850	& \multicolumn{2}{c}{2493}\\
				RF Stationary phase [$^o$]  						& 62.3 &58.9 & 63.0\\
				Momentum compaction,	$\alpha_c$ [$10^{-4}$]		& 1.3   &1.3& 0.88 \\
				Damping times, ($\tau_x, \tau_y, \tau_{l}$) [ms] 	&(1.98, 2.05, 1.04) &(1.24, 1.28, 0.33)&(1.19, 1.23, 0.31)\\
				Energy loss/turn, $U$ [MeV]							& 4.0  &5.6& 5.8 \\
				Quadrupole gradient strengths, ($k_1$, $k_2$) [T/m]	&(26, -53)	&(66, -98) &  (67, -98)\\
				Phase advances per arc cell, $(\mu_x, \mu_y)$ [$360^o$] & 0.408/0.050  & \multicolumn{2}{c}{0.510/0.110} \\
				Horizontal and vertical tune, ($Q_x$, $Q_y$) 	    &(48.35, 10.40) &(51.16, 14.56) & (51.18, 14.55)\\
				Horizontal and vertical chromaticity, ($\xi_x$, $\xi_y$) &(-113, -82) &(-57, -70) & (-67, -75)\\   
				\hline
				TME cell length, $L_{cell}$ [m]               	   &2.44& \multicolumn{2}{c}{2.62} \\
				Dipole field, ($B_{min}, B_{max}$) [T]		        & (0.97, 0.97) & (0.97, 0.97)  & (0.62, 2.32) \\
				Lengths and bending radii ratio, ($\lambda, \tilde{\rho}$) & (1, 1)   & (1, 1) & (0.04, 0.26) \\
				Normalized gradient in dipole  [m$^{-2}$ or T/m]   &-1.1 or -10.5	& 	\multicolumn{2}{c}{-1.1 or -10.5}  \\
				\hline      
				Wiggler peak field, $B_w$ [T]                       & 2.5  & \multicolumn{2}{c}{3.5}\\  
				Wiggler length, $L_{w}$ [m]                        &2 & \multicolumn{2}{c}{2} \\  
				Wiggler period, $\lambda_w$ [cm]					& 5.0  & \multicolumn{2}{c}{4.9} \\
				\hline
				& \multicolumn{2}{c}{without IBS} \\
				\hline		
				Normalized horiz. emittance, $\gamma\epsilon_{x}$ [nm-rad] & 312.2 &574.1 &  350.3 \\
				Normalized vert. emittance, $\gamma\epsilon_{y}$ [nm-rad]  & 3.3   &\multicolumn{2}{c}{3.3}    \\
				Energy spread (rms),	$\sigma_{\delta}$ [$\%$]		   & 0.11  &\multicolumn{2}{c}{0.15}\\
				Bunch length (rms), $\sigma_s$ [mm]                   	   & 1.4   &\multicolumn{2}{c}{1.6}\\
				Longitudinal emittance, $\epsilon_l$ [keVm] 			   & 4.4   &\multicolumn{2}{c}{5.8}\\
				\hline
				& \multicolumn{2}{c}{with IBS} \\
				\hline			
				Normalized horiz. emittance, $\gamma\epsilon_{x}$ [nm-rad]       & 478.9 &648.7  &434.7 \\
				Normalized vert. emittance, $\gamma\epsilon_{y}$ [nm-rad]        & 5.0   &4.5  &4.2    \\
				Energy spread (rms),	$\sigma_{\delta}$ [$\%$]		 		 & 0.11  &\multicolumn{2}{c}{0.15}\\
				Bunch length (rms), $\sigma_s$ [mm]                   			 & 1.5   &\multicolumn{2}{c}{1.6}\\
				Longitudinal emittance, $\epsilon_l$ [keVm] 			  		 & 4.7   &\multicolumn{2}{c}{5.8}\\
				IBS factors hor./ver./long. 		& 1.53/1.52/1.08 &1.13/1.35/1.01  &1.24/1.26/1.02\\
				\hline\hline
			\end{tabular}
		}
		\vspace{-10pt} 
	\end{center}
\end{table*}

\subsection{Dynamic aperture}
\begin{figure}[h!]
	\centering\includegraphics[width=.4\linewidth]{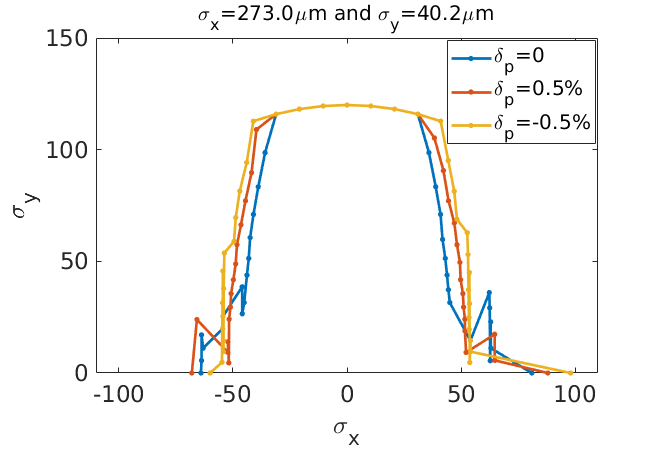}
	\captionsetup{justification=raggedright,singlelinecheck=false}
	\caption{{\small The on and off momentum dynamic aperture of the DR  for the trapezium dipole profile, for $\delta p=0$, $\delta p=0.5\%$ and $\delta p=-0.5\%$.}}
	\label{fig:da}
\end{figure}
\par The on- and off-momentum dynamic aperture (DA) of the ring was estimated for particles tracked with the PTC module of MADX~\cite{ref:madx}. Fig.~\ref{fig:da} shows the maximum initial positions of particles that survived over 1000 turns,  normalized to the horizontal and vertical beam sizes, at the point of calculation ($\sigma_x=273.0$ $\mu m$ and $\sigma_y=40.2$ $\mu m$). This simulation includes the effect of chromaticity sextupoles and magnets fringe fields but no other additional imperfection such as misalignments or magnet errors. The results for $\delta p=0$ are shown in blue, for $\delta p=0.5\%$ in red and for $\delta p=-0.5\%$ in yellow. The dynamic aperture achieved is remarkable (almost 14~mm in the horizontal plane and 5~mm in the vertical plane), allowing very comfortable on-axis injection. A working point optimization,  with simulations including misalignments, coupling and their correction, the non-linear effect of wigglers and space charge tune-shift, was further studied to fully quantify the non-linear performance of the new design~\cite{ref:Hossein}, which was found robust and adequate.

\section{Conclusions}
\par An analytical parametrization of the TME cell, based on linear optics considerations and the thin lens approximation, has been derived for the case of longitudinally variable bends. By choosing dipole profiles with longitudinally varying magnetic fields, subject to certain constraints, the emittances reached are lower as compared to the ones achieved by uniform dipoles. Among the non-uniform dipoles studied, it is found that the one having a bending radius forming a trapezium profile provides the largest emittance reduction. The design of the hyperbolic field profile that uses permanent magnets reaches  2.3~T at the highest field region and performs very well in terms of emittance reduction~\cite{ref:MagnetDesign2}. This innovative design could be applied in any low emittance ring.
\par The phase advances that determine the properties of a TME cell with the designed variable bend, can be chosen following the design goals. The applied optics stability and magnet feasibility requirements constrain the cell characteristics, starting from the drift lengths and the polarity of the quadrupoles. 
\par This analytical approach is used in order to define the appropriate initial conditions for matching the lattice and for finding optimal regions of operation for the best performance, subsequently verified numerically with MADX.  The CLIC DRs are chosen for numerically applying the concept of including variable bends in the TME arc cells, and thereby demonstrate clearly the reduction of the emittance as compared to the previous design, with uniform dipoles. 
The emittance is so much reduced and lower with respect to the target one, allowing the removal of a few TME cells. 
Moreover, by using Nb$_3$Sn high-field wigglers the final emittances including IBS are significantly reduced, as well, resulting to the elimination of a few LSS FODO cells. As the chromaticities are kept low, and the TME cell horizontal phase advance is close to $\pi$, the new proposed design achieves excellent DA that can be improved after a  further working point optimization~\cite{ref:Hossein}. The alternative DR design achieves all target parameters (even for increased bunch charge) although the circumference is reduced by around 13~\%, as compared to the original one.

\section{Acknowledgements}
We are particularly indebted to two anonymous referees whose constructive criticisms improved by far the clarity and quality of the paper. The research leading to these results has received funding from the European Commission under the FP7 Research Infrastructures project EuCARD-2, grant agreement no.312453 and the European Union’s Horizon 2020 Research and Innovation programme under Grant Agreement No 730871.
\appendix

\section{Quadrupole focal lengths in the limit of small drift lengths}
\label{focal}
Following Eqs.~\eqref{eq:f}, the full expressions of the quadrupole focal lengths, $f_1$ and $f_2$, at the limits where the drifts $s_1$, $s_2$ and $s_3$ are zeroed are:
\begin{itemize}
\item $\lim_{s_1\to 0} f_{1}=\dfrac{s_{2}(\eta_{cd} + \tilde{\theta})}{\eta_{cd}+ \tilde{\theta}-\eta_{ss1} + s_{2} \theta  } ~ \mathrm{and}~\lim_{s_1\to 0} f_{2} =\dfrac{s_{2} \eta_{ss1}}{\eta_{cd}+ \tilde{\theta}-\eta_{ss1}}$
\item $\lim_{s_2\to 0} f_{1}=0
	~~ \mathrm{and} ~~\lim_{s_2\to 0} f_{2} =0$
\item $\lim_{s_3\to 0} f_{1}=\dfrac{s_{2}g_1}
	{ g_1 + s_{2}  \theta  }
	~~ \mathrm{and}~~\lim_{s_3\to 0} f_{2} =0$ 
\end{itemize}
where: 
\begin{equation*}
\eta_{ss1}=\dfrac{\dfrac{-2(\eta_{cd}+ \tilde{\theta}) s_3}{s_2}}{1 \pm \sqrt{1 + \dfrac{4(\eta_{cd}+ \tilde{\theta}) s_3}{s_2^2} \dfrac{\beta_{cd}^2 \theta - g_2L/2 }{\beta_{cd}^2 \theta^2+g_2^2} }}
\end{equation*}
and for $g_1$ and $g_2$ given by  Eqs.~\eqref{eq:g}. 


\section{Integrals for the emittance of a variable bend}
\label{integrals}
In order to calculate the emittance of a variable bend described by two different bending radii functions along its length, Eq.~\eqref{eq:exelegant} is used, with the integrals $I_1$-$I_8$ being:
\begin{equation}
\label{eq:integ}
I_{1}=\mathlarger {\int\limits}_0^{L_{1}} \dfrac{ \theta_1^2 }{\left| \rho_1(s)  \right|^3 }ds 
\hspace{0.1cm1} , \hspace{0.2cm}
I_{2}=\mathlarger {\int\limits}_{L_{1}}^{L_{1}+L_{2}} \dfrac{ (\theta_2+\theta_{L_1})^2 }{\left| \rho_2(s)  \right|^3 }ds
\hspace{0.1cm} , \hspace{0.2cm}
I_{3}=\mathlarger {\int\limits}_0^{L_{1}} \dfrac{1}{\left| \rho_1(s)  \right|^3 }ds 
\hspace{0.1cm} , \hspace{0.2cm} I_{4}=\mathlarger {\int\limits}_{L_{1}}^{L_{1}+L_{2}} \dfrac{1}{\left| \rho_2(s)  \right|^3 }ds
\hspace{0.1cm} , \hspace{0.2cm}
I_{5}=\mathlarger {\int\limits}_0^{L_{1}} 2\dfrac{-s \theta_1 + \tilde{\theta}_1}{\left| \rho_1(s)  \right|^3 }ds  
\end{equation}
\begin{equation*}
I_{6}=\mathlarger {\int\limits}_{L_{1}}^{L_{1}+L_{2}}
2\dfrac{-s \theta_2 + \tilde{\theta}_2-L_1 \theta_{L_1}+\tilde{\theta}_{L_1}}{\left| \rho_2(s)  \right|^3 }ds 
\hspace{0.1cm} , \hspace{0.2cm}
I_{7}=\mathlarger {\int\limits}_0^{L_{1}} \dfrac{(-s  \theta_1 + \tilde{\theta}_1)^2}{\left| \rho_1(s)  \right|^3 }ds 
\hspace{0.1cm} , \hspace{0.2cm}
I_{8}=\mathlarger{\int\limits}_{L_{1}}^{L_{1}+L_{2}}
\dfrac{(-s  \theta_2 + \tilde{\theta}_2-L_1  \theta_{L_{1}} + \tilde{\theta}_{L_{1}})^2}{\left| \rho_2(s)  \right|^3 }ds\;\;,
\end{equation*}
where $\theta_{L_{1}}=\theta(s=L_1)$ and $\tilde{\theta}_{L_{1}}=\tilde{\theta}(s=L_1)$.

\section{Emittance reduction factor}
\label{redfactor}
As discussed in section~\ref{dipprof}, the emittance reduction factors for the step and the trapezium variable bend profiles depend only on $\tilde{\rho}$ and $\lambda$ and their full expressions are given by: 

\begin{equation*}
F_{TME_{step}}=2 (\lambda + \tilde{\rho})^3 (1 + \tilde{\rho}^2)\sqrt{\dfrac{1 + \tilde{\rho}^3}
{\lambda_1\tilde{\rho}^4+\lambda_2\tilde{\rho}^5+\lambda_3\tilde{\rho}^7+\lambda_4\tilde{\rho}^8+\lambda_5 \tilde{\rho}^9+\lambda_6\tilde{\rho}^{10}+\lambda_7 (1+9 \tilde{\rho}^3+18 \tilde{\rho}^6)+\lambda_8 (\lambda_9 \tilde{\rho}^{11}+\lambda_{10} \tilde{\rho}^{12}+\tilde{\rho}^{13})}}~,
\end{equation*}
where $\lambda_1= 4 \lambda^4 (5+18 \lambda)$,
$\lambda_2=\lambda^2 (9+45 \lambda+64 \lambda^2)$,
$\lambda_3=12 \lambda^4 (5+21 \lambda)$,
$\lambda_4=\lambda^2 (31+210 \lambda+399 \lambda^2)$,\\
$\lambda_5=\lambda (27+155 \lambda+240 \lambda^2)$,
$\lambda_6=3 (3+15 \lambda+20 \lambda^2)$,
$\lambda_7=4 \lambda^6$,
$\lambda_8=(4+15 \lambda+15 \lambda^2)$,
$\lambda_9=3\lambda_7/(4\lambda^4) $ and 
$\lambda_{10}=\lambda_9/ \lambda$~.

\begin{equation*}
F_{TME_{trapezium}}=\dfrac{4 \sqrt{2}r_1 (r_2 \lambda + \tilde{\rho} \ln\tilde{\rho})^3 \sqrt{w_2/w_3}}{w_1}~,
\end{equation*}
where $r_1=\tilde{\rho}+1$, $r_2=\tilde{\rho}-1$ and:
\begin{equation*}
w_1=2r_2^3(2+3r_1\tilde{\rho})\lambda^2-6r_2\tilde{\rho}^2(r_1r_2-2\tilde{\rho}^2\ln\tilde{\rho})\lambda+3(r_1r_2-2\tilde{\rho}^2\ln\tilde{\rho})\tilde{\rho}^3+6\tilde{\rho}^5(\ln\tilde{\rho})^2~,
\end{equation*}
\begin{equation*}
w_2=r_2^3(2+r_1\tilde{\rho})w_1~,
\end{equation*}
\begin{equation*}
w_3=w_{3a}\lambda+w_{3b}\lambda^2+w_{3c}\lambda^3+w_{3d}\lambda^4+w_{3e}~,
\end{equation*}
for 
$
w_{3a}=90r_2\tilde{\rho}^4(r_2^2 (-8 + \tilde{\rho} (5 + (-1 + r_2) \tilde{\rho})) + 
4 (-1 + r_2) \tilde{\rho} \ln\tilde{\rho} (-2 r_2 + \tilde{\rho} \ln\tilde{\rho})),\\
w_{3b}=15r_2^2\tilde{\rho}^3(r_2^2 (20 + 3 (2 + r_1) r_2 \tilde{\rho}) + 4 \tilde{\rho}^2 \ln\tilde{\rho} (-2 r_2 + 3 (-1 + r_2) \ln\tilde{\rho})),
w_{3c}=-120r_2^4\tilde{\rho}^2(r_1r_2-2\tilde{\rho}^2\ln\tilde{\rho}),\\
w_{3d}=16r_2^6(1+3r_1\tilde{\rho})~\mathrm{and}~
w_{3e}=45 \tilde{\rho}^5 (r_2^2 (-44 + \tilde{\rho} (-3 + (-5 + r_2) \tilde{\rho})) + 
4 \ln\tilde{\rho} (4 r_1 r_2 + 
6 r_2 \tilde{\rho} + (-1 + r_2) \tilde{\rho}^2 \ln\tilde{\rho}))$~.

\bibliography{clic_ref}

\begin{thebibliography}{25}%
\makeatletter
\providecommand \@ifxundefined [1]{%
 \@ifx{#1\undefined}
}%
\providecommand \@ifnum [1]{%
 \ifnum #1\expandafter \@firstoftwo
 \else \expandafter \@secondoftwo
 \fi
}%
\providecommand \@ifx [1]{%
 \ifx #1\expandafter \@firstoftwo
 \else \expandafter \@secondoftwo
 \fi
}%
\providecommand \natexlab [1]{#1}%
\providecommand \enquote  [1]{``#1''}%
\providecommand \bibnamefont  [1]{#1}%
\providecommand \bibfnamefont [1]{#1}%
\providecommand \citenamefont [1]{#1}%
\providecommand \href@noop [0]{\@secondoftwo}%
\providecommand \href [0]{\begingroup \@sanitize@url \@href}%
\providecommand \@href[1]{\@@startlink{#1}\@@href}%
\providecommand \@@href[1]{\endgroup#1\@@endlink}%
\providecommand \@sanitize@url [0]{\catcode `\\12\catcode `\$12\catcode
  `\&12\catcode `\#12\catcode `\^12\catcode `\_12\catcode `\%12\relax}%
\providecommand \@@startlink[1]{}%
\providecommand \@@endlink[0]{}%
\providecommand \url  [0]{\begingroup\@sanitize@url \@url }%
\providecommand \@url [1]{\endgroup\@href {#1}{\urlprefix }}%
\providecommand \urlprefix  [0]{URL }%
\providecommand \Eprint [0]{\href }%
\providecommand \doibase [0]{http://dx.doi.org/}%
\providecommand \selectlanguage [0]{\@gobble}%
\providecommand \bibinfo  [0]{\@secondoftwo}%
\providecommand \bibfield  [0]{\@secondoftwo}%
\providecommand \translation [1]{[#1]}%
\providecommand \BibitemOpen [0]{}%
\providecommand \bibitemStop [0]{}%
\providecommand \bibitemNoStop [0]{.\EOS\space}%
\providecommand \EOS [0]{\spacefactor3000\relax}%
\providecommand \BibitemShut  [1]{\csname bibitem#1\endcsname}%
\let\auto@bib@innerbib\@empty
\bibitem [{\citenamefont {Teng}(1985)}]{ref:tme0}%
  \BibitemOpen
  \bibfield  {author} {\bibinfo {author} {\bibfnamefont {L.~C.}\ \bibnamefont
  {Teng}},\ }\href@noop {} {\enquote {\bibinfo {title} {Minimum emittance
  lattice for synchrotron radiation storage rings},}\ } (\bibinfo {year}
  {1985})\BibitemShut {NoStop}%
\bibitem [{\citenamefont {Einfeld}\ \emph {et~al.}(1996)\citenamefont
  {Einfeld}, \citenamefont {Schaper},\ and\ \citenamefont {Plesko}}]{ref:tme}%
  \BibitemOpen
  \bibfield  {author} {\bibinfo {author} {\bibfnamefont {D.}~\bibnamefont
  {Einfeld}}, \bibinfo {author} {\bibfnamefont {J.}~\bibnamefont {Schaper}}, \
  and\ \bibinfo {author} {\bibfnamefont {M.}~\bibnamefont {Plesko}},\
  }\href@noop {} {\enquote {\bibinfo {title} {A lattice design to reach the
  theoretical minimum emittance for a storage ring},}\ } (\bibinfo {year}
  {1996})\BibitemShut {NoStop}%
\bibitem [{\citenamefont {Antoniou}\ and\ \citenamefont
  {Papaphilippou}(2014)}]{ref:fan}%
  \BibitemOpen
  \bibfield  {author} {\bibinfo {author} {\bibfnamefont {F.}~\bibnamefont
  {Antoniou}}\ and\ \bibinfo {author} {\bibfnamefont {Y.}~\bibnamefont
  {Papaphilippou}},\ }\href {\doibase 10.1103/PhysRevSTAB.17.064002} {\bibfield
   {journal} {\bibinfo  {journal} {Phys. Rev. ST Accel. Beams}\ }\textbf
  {\bibinfo {volume} {17}},\ \bibinfo {pages} {064002} (\bibinfo {year}
  {2014})}\BibitemShut {NoStop}%
\bibitem [{\citenamefont {Guo}\ and\ \citenamefont
  {Raubenheimer}(2002)}]{ref:gr}%
  \BibitemOpen
  \bibfield  {author} {\bibinfo {author} {\bibfnamefont {J.}~\bibnamefont
  {Guo}}\ and\ \bibinfo {author} {\bibfnamefont {T.}~\bibnamefont
  {Raubenheimer}},\ }in\ \href
  {http://accelconf.web.cern.ch/AccelConf/e02/PAPERS/WEPLE001.pdf} {\emph
  {\bibinfo {booktitle} {{Particle accelerator. Proceedings, 8th European
  Conference, EPAC 2002, Paris, France, June 3-7, 2002}}}}\ (\bibinfo {year}
  {2002})\ pp.\ \bibinfo {pages} {1136--1138}\BibitemShut {NoStop}%
\bibitem [{\citenamefont {Nagaoka}\ and\ \citenamefont
  {Wrulich}(2007)}]{ref:nw}%
  \BibitemOpen
  \bibfield  {author} {\bibinfo {author} {\bibfnamefont {R.}~\bibnamefont
  {Nagaoka}}\ and\ \bibinfo {author} {\bibfnamefont {A.~F.}\ \bibnamefont
  {Wrulich}},\ }\href {\doibase https://doi.org/10.1016/j.nima.2007.02.086}
  {\bibfield  {journal} {\bibinfo  {journal} {Nuclear Instruments and Methods
  in Physics Research Section A: Accelerators, Spectrometers, Detectors and
  Associated Equipment}\ }\textbf {\bibinfo {volume} {575}},\ \bibinfo {pages}
  {292 } (\bibinfo {year} {2007})}\BibitemShut {NoStop}%
\bibitem [{\citenamefont {{MAD-X homepage}}()}]{ref:madx}%
  \BibitemOpen
  \bibfield  {author} {\bibinfo {author} {\bibnamefont {{MAD-X homepage}}},\
  }\href@noop {} {}\bibinfo {howpublished} {URL
  \url{http://http://mad.web.cern.ch/mad/}}\BibitemShut {NoStop}%
\bibitem [{\citenamefont {Papaphilippou}\ and\ \citenamefont
  {Elleaume}(2005)}]{ref:pe}%
  \BibitemOpen
  \bibfield  {author} {\bibinfo {author} {\bibfnamefont {Y.}~\bibnamefont
  {Papaphilippou}}\ and\ \bibinfo {author} {\bibfnamefont {P.}~\bibnamefont
  {Elleaume}},\ }in\ \href {\doibase 10.1109/PAC.2005.1591017} {\emph {\bibinfo
  {booktitle} {Proceedings of the 2005 Particle Accelerator Conference}}}\
  (\bibinfo {year} {2005})\ pp.\ \bibinfo {pages} {2086--2088}\BibitemShut
  {NoStop}%
\bibitem [{\citenamefont {Papaphilippou}\ \emph {et~al.}(2005)\citenamefont
  {Papaphilippou}, \citenamefont {Ropert}, \citenamefont {Elleaume},\ and\
  \citenamefont {Farvacque}}]{ref:esrf}%
  \BibitemOpen
  \bibfield  {author} {\bibinfo {author} {\bibfnamefont {Y.}~\bibnamefont
  {Papaphilippou}}, \bibinfo {author} {\bibfnamefont {A.}~\bibnamefont
  {Ropert}}, \bibinfo {author} {\bibfnamefont {P.}~\bibnamefont {Elleaume}}, \
  and\ \bibinfo {author} {\bibfnamefont {L.}~\bibnamefont {Farvacque}},\ }in\
  \href {\doibase 10.1109/PAC.2005.1591004} {\emph {\bibinfo {booktitle}
  {Proceedings of the 2005 Particle Accelerator Conference}}}\ (\bibinfo {year}
  {2005})\ pp.\ \bibinfo {pages} {2047--2049}\BibitemShut {NoStop}%
\bibitem [{\citenamefont {Wang}\ \emph {et~al.}(2011)\citenamefont {Wang},
  \citenamefont {Wang},\ and\ \citenamefont {Peng}}]{ref:wwp}%
  \BibitemOpen
  \bibfield  {author} {\bibinfo {author} {\bibfnamefont {C.-x.}\ \bibnamefont
  {Wang}}, \bibinfo {author} {\bibfnamefont {Y.}~\bibnamefont {Wang}}, \ and\
  \bibinfo {author} {\bibfnamefont {Y.}~\bibnamefont {Peng}},\ }\href {\doibase
  10.1103/PhysRevSTAB.14.034001} {\bibfield  {journal} {\bibinfo  {journal}
  {Phys. Rev. ST Accel. Beams}\ }\textbf {\bibinfo {volume} {14}},\ \bibinfo
  {pages} {034001} (\bibinfo {year} {2011})}\BibitemShut {NoStop}%
\bibitem [{\citenamefont {Wang}(2009)}]{ref:wang}%
  \BibitemOpen
  \bibfield  {author} {\bibinfo {author} {\bibfnamefont {C.-x.}\ \bibnamefont
  {Wang}},\ }\href {\doibase 10.1103/PhysRevSTAB.12.061001} {\bibfield
  {journal} {\bibinfo  {journal} {Phys. Rev. ST Accel. Beams}\ }\textbf
  {\bibinfo {volume} {12}},\ \bibinfo {pages} {061001} (\bibinfo {year}
  {2009})}\BibitemShut {NoStop}%
\bibitem [{\citenamefont {Papadopoulou}()}]{ref:ler14_stef}%
  \BibitemOpen
  \bibfield  {author} {\bibinfo {author} {\bibfnamefont {S.}~\bibnamefont
  {Papadopoulou}},\ }\href
  {https://agenda.infn.it/getFile.py/access?contribId=23&sessionId=0&resId=0&materialId=slides&confId=7766}
  {\enquote {\bibinfo {title} {Analytical considerations for reducing the
  emittance with longitudinally variable bends},}\ }\bibinfo {note} {{presented
  at the 4th Low Emittance Rings Workshop (LOWεRING 2014), 17-19 September
  2014, INFN-LNF, Frascati, Italy}}\BibitemShut {NoStop}%
\bibitem [{\citenamefont {Streun}\ and\ \citenamefont
  {Wrulich}(2015)}]{ref:streun15}%
  \BibitemOpen
  \bibfield  {author} {\bibinfo {author} {\bibfnamefont {A.}~\bibnamefont
  {Streun}}\ and\ \bibinfo {author} {\bibfnamefont {A.}~\bibnamefont
  {Wrulich}},\ }\href {\doibase https://doi.org/10.1016/j.nima.2014.10.002}
  {\bibfield  {journal} {\bibinfo  {journal} {Nuclear Instruments and Methods
  in Physics Research Section A: Accelerators, Spectrometers, Detectors and
  Associated Equipment}\ }\textbf {\bibinfo {volume} {770}},\ \bibinfo {pages}
  {98 } (\bibinfo {year} {2015})}\BibitemShut {NoStop}%
\bibitem [{\citenamefont {Papadopoulou}\ \emph {et~al.}(2015)\citenamefont
  {Papadopoulou}, \citenamefont {Antoniou},\ and\ \citenamefont
  {Papaphilippou}}]{ref:ipac15_stef}%
  \BibitemOpen
  \bibfield  {author} {\bibinfo {author} {\bibfnamefont {S.}~\bibnamefont
  {Papadopoulou}}, \bibinfo {author} {\bibfnamefont {F.}~\bibnamefont
  {Antoniou}}, \ and\ \bibinfo {author} {\bibfnamefont {Y.}~\bibnamefont
  {Papaphilippou}},\ }in\ \href
  {http://accelconf.web.cern.ch/AccelConf/IPAC2015/papers/tupty022.pdf} {\emph
  {\bibinfo {booktitle} {{Proceedings, 6th International Particle Accelerator
  Conference (IPAC 2015): Richmond, Virginia, USA, May 3-8, 2015}}}}\ (\bibinfo
  {year} {2015})\ p.\ \bibinfo {pages} {TUPTY022}\BibitemShut {NoStop}%
\bibitem [{\citenamefont {Liuzzo}\ \emph {et~al.}(2016)\citenamefont {Liuzzo},
  \citenamefont {Einfeld}, \citenamefont {Farvacque},\ and\ \citenamefont
  {Raimondi}}]{ref:esrf2}%
  \BibitemOpen
  \bibfield  {author} {\bibinfo {author} {\bibfnamefont {S.}~\bibnamefont
  {Liuzzo}}, \bibinfo {author} {\bibfnamefont {D.}~\bibnamefont {Einfeld}},
  \bibinfo {author} {\bibfnamefont {L.}~\bibnamefont {Farvacque}}, \ and\
  \bibinfo {author} {\bibfnamefont {P.}~\bibnamefont {Raimondi}},\ }in\ \href
  {\doibase 10.18429/JACoW-IPAC2016-WEPOW006} {\emph {\bibinfo {booktitle}
  {{Proceedings, 7th International Particle Accelerator Conference (IPAC 2016):
  Busan, Korea, May 8-13, 2016}}}}\ (\bibinfo {year} {2016})\ p.\ \bibinfo
  {pages} {WEPOW006}\BibitemShut {NoStop}%
\bibitem [{\citenamefont {Streun}(2006)}]{ref:streun}%
  \BibitemOpen
  \bibfield  {author} {\bibinfo {author} {\bibfnamefont {A.}~\bibnamefont
  {Streun}},\ }\href {https://cds.cern.ch/record/941327} {\emph {\bibinfo
  {title} {{Lattices for light sources}}}},\ \bibinfo {type} {Tech. Rep.}\
  \bibinfo {number} {CERN-2006-002}\ (\bibinfo {year} {2006})\BibitemShut
  {NoStop}%
\bibitem [{\citenamefont {Dom\'{i}nguez~Martinez}\ and\ \citenamefont
  {Toral}()}]{ref:MagnetDesign1}%
  \BibitemOpen
  \bibfield  {author} {\bibinfo {author} {\bibfnamefont {M.~A.}\ \bibnamefont
  {Dom\'{i}nguez~Martinez}}\ and\ \bibinfo {author} {\bibfnamefont
  {F.}~\bibnamefont {Toral}},\ }\href@noop {} {}\bibinfo {howpublished}
  {(private communication)}\BibitemShut {NoStop}%
\bibitem [{\citenamefont {Dom\'{i}nguez~Martinez}\ \emph
  {et~al.}(2018)\citenamefont {Dom\'{i}nguez~Martinez}, \citenamefont {Toral},
  \citenamefont {Ghasem}, \citenamefont {Papadopopoulou},\ and\ \citenamefont
  {Papaphilippou}}]{ref:MagnetDesign2}%
  \BibitemOpen
  \bibfield  {author} {\bibinfo {author} {\bibfnamefont {M.~A.}\ \bibnamefont
  {Dom\'{i}nguez~Martinez}}, \bibinfo {author} {\bibfnamefont {F.}~\bibnamefont
  {Toral}}, \bibinfo {author} {\bibfnamefont {H.}~\bibnamefont {Ghasem}},
  \bibinfo {author} {\bibfnamefont {P.~S.}\ \bibnamefont {Papadopopoulou}}, \
  and\ \bibinfo {author} {\bibfnamefont {Y.}~\bibnamefont {Papaphilippou}},\
  }\href {\doibase 10.1109/TASC.2018.2795551} {\bibfield  {journal} {\bibinfo
  {journal} {IEEE Transactions on Applied Superconductivity}\ }\textbf
  {\bibinfo {volume} {PP}},\ \bibinfo {pages} {1} (\bibinfo {year}
  {2018})}\BibitemShut {NoStop}%
\bibitem [{\citenamefont {Papaphilippou}\ \emph {et~al.}(2012)\citenamefont
  {Papaphilippou}, \citenamefont {Antoniou}, \citenamefont {Barnes},
  \citenamefont {Calatroni}, \citenamefont {Chiggiato}, \citenamefont
  {Corsini}, \citenamefont {Grudiev}, \citenamefont {Koukovini}, \citenamefont
  {Lefevre}, \citenamefont {Martini}, \citenamefont {Modena}, \citenamefont
  {Mounet}, \citenamefont {Perin}, \citenamefont {Renier}, \citenamefont
  {Russenschuck}, \citenamefont {Rumolo}, \citenamefont {Schoerling},
  \citenamefont {Schulte}, \citenamefont {Schmickler}, \citenamefont
  {Taborelli}, \citenamefont {Vandoni}, \citenamefont {Zimmermann},
  \citenamefont {Zisopoulos}, \citenamefont {Boland}, \citenamefont {Palmer},
  \citenamefont {Bragin}, \citenamefont {Levichev}, \citenamefont {Syniatkin},
  \citenamefont {Zolotarev}, \citenamefont {Vobly}, \citenamefont {Korostelev},
  \citenamefont {Vivoli}, \citenamefont {Belver-Aguilar}, \citenamefont
  {Faus-Golfe}, \citenamefont {Rinolfi}, \citenamefont {Bernhard},
  \citenamefont {Pivi}, \citenamefont {Smith}, \citenamefont {Rassool},\ and\
  \citenamefont {Wootton}}]{ref:clic}%
  \BibitemOpen
  \bibfield  {author} {\bibinfo {author} {\bibfnamefont {Y.}~\bibnamefont
  {Papaphilippou}}, \bibinfo {author} {\bibfnamefont {F.}~\bibnamefont
  {Antoniou}}, \bibinfo {author} {\bibfnamefont {M.}~\bibnamefont {Barnes}},
  \bibinfo {author} {\bibfnamefont {S.}~\bibnamefont {Calatroni}}, \bibinfo
  {author} {\bibfnamefont {P.}~\bibnamefont {Chiggiato}}, \bibinfo {author}
  {\bibfnamefont {R.}~\bibnamefont {Corsini}}, \bibinfo {author} {\bibfnamefont
  {A.}~\bibnamefont {Grudiev}}, \bibinfo {author} {\bibfnamefont
  {E.}~\bibnamefont {Koukovini}}, \bibinfo {author} {\bibfnamefont
  {T.}~\bibnamefont {Lefevre}}, \bibinfo {author} {\bibfnamefont
  {M.}~\bibnamefont {Martini}}, \bibinfo {author} {\bibfnamefont
  {M.}~\bibnamefont {Modena}}, \bibinfo {author} {\bibfnamefont
  {N.}~\bibnamefont {Mounet}}, \bibinfo {author} {\bibfnamefont
  {A.}~\bibnamefont {Perin}}, \bibinfo {author} {\bibfnamefont
  {Y.}~\bibnamefont {Renier}}, \bibinfo {author} {\bibfnamefont
  {S.}~\bibnamefont {Russenschuck}}, \bibinfo {author} {\bibfnamefont
  {G.}~\bibnamefont {Rumolo}}, \bibinfo {author} {\bibfnamefont
  {D.}~\bibnamefont {Schoerling}}, \bibinfo {author} {\bibfnamefont
  {D.}~\bibnamefont {Schulte}}, \bibinfo {author} {\bibfnamefont
  {H.}~\bibnamefont {Schmickler}}, \bibinfo {author} {\bibfnamefont
  {M.}~\bibnamefont {Taborelli}}, \bibinfo {author} {\bibfnamefont
  {G.}~\bibnamefont {Vandoni}}, \bibinfo {author} {\bibfnamefont
  {F.}~\bibnamefont {Zimmermann}}, \bibinfo {author} {\bibfnamefont
  {P.}~\bibnamefont {Zisopoulos}}, \bibinfo {author} {\bibfnamefont
  {M.}~\bibnamefont {Boland}}, \bibinfo {author} {\bibfnamefont
  {M.}~\bibnamefont {Palmer}}, \bibinfo {author} {\bibfnamefont
  {A.}~\bibnamefont {Bragin}}, \bibinfo {author} {\bibfnamefont
  {E.}~\bibnamefont {Levichev}}, \bibinfo {author} {\bibfnamefont
  {S.}~\bibnamefont {Syniatkin}}, \bibinfo {author} {\bibfnamefont
  {K.}~\bibnamefont {Zolotarev}}, \bibinfo {author} {\bibfnamefont
  {P.}~\bibnamefont {Vobly}}, \bibinfo {author} {\bibfnamefont
  {M.}~\bibnamefont {Korostelev}}, \bibinfo {author} {\bibfnamefont
  {A.}~\bibnamefont {Vivoli}}, \bibinfo {author} {\bibfnamefont
  {C.}~\bibnamefont {Belver-Aguilar}}, \bibinfo {author} {\bibfnamefont
  {A.}~\bibnamefont {Faus-Golfe}}, \bibinfo {author} {\bibfnamefont
  {L.}~\bibnamefont {Rinolfi}}, \bibinfo {author} {\bibfnamefont
  {A.}~\bibnamefont {Bernhard}}, \bibinfo {author} {\bibfnamefont
  {M.}~\bibnamefont {Pivi}}, \bibinfo {author} {\bibfnamefont {S.}~\bibnamefont
  {Smith}}, \bibinfo {author} {\bibfnamefont {R.}~\bibnamefont {Rassool}}, \
  and\ \bibinfo {author} {\bibfnamefont {K.}~\bibnamefont {Wootton}},\
  }\href@noop {} {\emph {\bibinfo {title} {Conceptual Design of the CLIC
  Damping Rings.}}},\ \bibinfo {type} {Tech. Rep.}\ \bibinfo {number}
  {CERN-ATS-2012-176}\ (\bibinfo  {institution} {CERN},\ \bibinfo {address}
  {Geneva},\ \bibinfo {year} {2012})\BibitemShut {NoStop}%
\bibitem [{\citenamefont {Jiao}\ \emph {et~al.}(2011)\citenamefont {Jiao},
  \citenamefont {Cai},\ and\ \citenamefont {Chao}}]{ref:modified}%
  \BibitemOpen
  \bibfield  {author} {\bibinfo {author} {\bibfnamefont {Y.}~\bibnamefont
  {Jiao}}, \bibinfo {author} {\bibfnamefont {Y.}~\bibnamefont {Cai}}, \ and\
  \bibinfo {author} {\bibfnamefont {A.~W.}\ \bibnamefont {Chao}},\ }\href
  {\doibase 10.1103/PhysRevSTAB.14.054002} {\bibfield  {journal} {\bibinfo
  {journal} {Phys. Rev. ST Accel. Beams}\ }\textbf {\bibinfo {volume} {14}},\
  \bibinfo {pages} {054002} (\bibinfo {year} {2011})}\BibitemShut {NoStop}%
\bibitem [{\citenamefont {Braun}\ \emph {et~al.}(2010)\citenamefont {Braun},
  \citenamefont {Levichev}, \citenamefont {Papaphilippou}, \citenamefont
  {Siniatkin},\ and\ \citenamefont {Zolotarev}}]{ref:grad}%
  \BibitemOpen
  \bibfield  {author} {\bibinfo {author} {\bibfnamefont {H.}~\bibnamefont
  {Braun}}, \bibinfo {author} {\bibfnamefont {E.}~\bibnamefont {Levichev}},
  \bibinfo {author} {\bibfnamefont {Y.}~\bibnamefont {Papaphilippou}}, \bibinfo
  {author} {\bibfnamefont {S.}~\bibnamefont {Siniatkin}}, \ and\ \bibinfo
  {author} {\bibfnamefont {K.}~\bibnamefont {Zolotarev}},\ }\href
  {https://cds.cern.ch/record/1347751} {\emph {\bibinfo {title} {{Alternative
  design of the CLIC Damping Ring Lattice}}}},\ \bibinfo {type} {Tech. Rep.}\
  \bibinfo {number} {CERN-OPEN-2011-016. CLIC-Note-849}\ (\bibinfo
  {institution} {BINP},\ \bibinfo {address} {Novosibirsk},\ \bibinfo {year}
  {2010})\BibitemShut {NoStop}%
\bibitem [{\citenamefont {Antoniou}(2012)}]{ref:thesis}%
  \BibitemOpen
  \bibfield  {author} {\bibinfo {author} {\bibfnamefont {F.}~\bibnamefont
  {Antoniou}},\ }\emph {\bibinfo {title} {{Optics design of Intrabeam
  Scattering dominated damping rings}}},\ \href
  {http://cds.cern.ch/record/1666863} {Ph.D. thesis} (\bibinfo {year} {2012}),\
  \bibinfo {note} {defended on 08 Jan 2013}\BibitemShut {NoStop}%
\bibitem [{\citenamefont {Schoerling}\ \emph {et~al.}(2012)\citenamefont
  {Schoerling}, \citenamefont {Antoniou}, \citenamefont {Bernhard},
  \citenamefont {Bragin}, \citenamefont {Karppinen}, \citenamefont
  {Maccaferri}, \citenamefont {Mezentsev}, \citenamefont {Papaphilippou},
  \citenamefont {Peiffer}, \citenamefont {Rossmanith}, \citenamefont {Rumolo},
  \citenamefont {Russenschuck}, \citenamefont {Vobly},\ and\ \citenamefont
  {Zolotarev}}]{ref:wiggler}%
  \BibitemOpen
  \bibfield  {author} {\bibinfo {author} {\bibfnamefont {D.}~\bibnamefont
  {Schoerling}}, \bibinfo {author} {\bibfnamefont {F.}~\bibnamefont
  {Antoniou}}, \bibinfo {author} {\bibfnamefont {A.}~\bibnamefont {Bernhard}},
  \bibinfo {author} {\bibfnamefont {A.}~\bibnamefont {Bragin}}, \bibinfo
  {author} {\bibfnamefont {M.}~\bibnamefont {Karppinen}}, \bibinfo {author}
  {\bibfnamefont {R.}~\bibnamefont {Maccaferri}}, \bibinfo {author}
  {\bibfnamefont {N.}~\bibnamefont {Mezentsev}}, \bibinfo {author}
  {\bibfnamefont {Y.}~\bibnamefont {Papaphilippou}}, \bibinfo {author}
  {\bibfnamefont {P.}~\bibnamefont {Peiffer}}, \bibinfo {author} {\bibfnamefont
  {R.}~\bibnamefont {Rossmanith}}, \bibinfo {author} {\bibfnamefont
  {G.}~\bibnamefont {Rumolo}}, \bibinfo {author} {\bibfnamefont
  {S.}~\bibnamefont {Russenschuck}}, \bibinfo {author} {\bibfnamefont
  {P.}~\bibnamefont {Vobly}}, \ and\ \bibinfo {author} {\bibfnamefont
  {K.}~\bibnamefont {Zolotarev}},\ }\href {\doibase
  10.1103/PhysRevSTAB.15.042401} {\bibfield  {journal} {\bibinfo  {journal}
  {Phys. Rev. ST Accel. Beams}\ }\textbf {\bibinfo {volume} {15}},\ \bibinfo
  {pages} {042401} (\bibinfo {year} {2012})}\BibitemShut {NoStop}%
\bibitem [{\citenamefont {Garc\'{i}a~Fajardo}\ \emph
  {et~al.}(2016)\citenamefont {Garc\'{i}a~Fajardo}, \citenamefont {Antoniou},
  \citenamefont {Bernhard}, \citenamefont {Ferracin}, \citenamefont {Mazet},
  \citenamefont {Papadopoulou}, \citenamefont {Papaphilippou}, \citenamefont
  {Pérez},\ and\ \citenamefont {Schoerling}}]{ref:laura}%
  \BibitemOpen
  \bibfield  {author} {\bibinfo {author} {\bibfnamefont {L.}~\bibnamefont
  {Garc\'{i}a~Fajardo}}, \bibinfo {author} {\bibfnamefont {F.}~\bibnamefont
  {Antoniou}}, \bibinfo {author} {\bibfnamefont {A.}~\bibnamefont {Bernhard}},
  \bibinfo {author} {\bibfnamefont {P.}~\bibnamefont {Ferracin}}, \bibinfo
  {author} {\bibfnamefont {J.}~\bibnamefont {Mazet}}, \bibinfo {author}
  {\bibfnamefont {S.}~\bibnamefont {Papadopoulou}}, \bibinfo {author}
  {\bibfnamefont {Y.}~\bibnamefont {Papaphilippou}}, \bibinfo {author}
  {\bibfnamefont {J.~C.}\ \bibnamefont {Pérez}}, \ and\ \bibinfo {author}
  {\bibfnamefont {D.}~\bibnamefont {Schoerling}},\ }\bibfield  {booktitle}
  {\emph {\bibinfo {booktitle} {{Proceedings, 24th International Conference on
  Magnet Technology (MT-24): Seoul, Korea, October 18-23, 2015}}},\ }\href
  {\doibase 10.1109/TASC.2016.2517938} {\bibfield  {journal} {\bibinfo
  {journal} {IEEE Trans. Appl. Supercond.}\ }\textbf {\bibinfo {volume} {26}},\
  \bibinfo {pages} {4100506} (\bibinfo {year} {2016})}\BibitemShut {NoStop}%
\bibitem [{\citenamefont {Boland}\ \emph {et~al.}(2016)\citenamefont {Boland}
  \emph {et~al.}}]{ref:bol}%
  \BibitemOpen
  \bibfield  {author} {\bibinfo {author} {\bibfnamefont {M.~J.}\ \bibnamefont
  {Boland}} \emph {et~al.} (\bibinfo {collaboration} {CLIC and CLICdp
  collaborations}),\ }\href {https://cds.cern.ch/record/2210892} {\emph
  {\bibinfo {title} {{Updated baseline for a staged Compact Linear
  Collider}}}},\ \bibinfo {type} {Tech. Rep.}\ \bibinfo {number}
  {CERN-2016-004}\ (\bibinfo {address} {Geneva},\ \bibinfo {year}
  {2016})\BibitemShut {NoStop}%
\bibitem [{\citenamefont {Ghasem}\ \emph {et~al.}(2016)\citenamefont {Ghasem},
  \citenamefont {Alabau-Gonzalvo}, \citenamefont {Antoniou}, \citenamefont
  {Papadopoulou},\ and\ \citenamefont {Papaphilippou}}]{ref:Hossein}%
  \BibitemOpen
  \bibfield  {author} {\bibinfo {author} {\bibfnamefont {H.}~\bibnamefont
  {Ghasem}}, \bibinfo {author} {\bibfnamefont {J.}~\bibnamefont
  {Alabau-Gonzalvo}}, \bibinfo {author} {\bibfnamefont {F.}~\bibnamefont
  {Antoniou}}, \bibinfo {author} {\bibfnamefont {S.}~\bibnamefont
  {Papadopoulou}}, \ and\ \bibinfo {author} {\bibfnamefont {Y.}~\bibnamefont
  {Papaphilippou}},\ }in\ \href {\doibase 10.18429/JACoW-IPAC2016-WEPMW003}
  {\emph {\bibinfo {booktitle} {{Proceedings, 7th International Particle
  Accelerator Conference (IPAC 2016): Busan, Korea, May 8-13, 2016}}}}\
  (\bibinfo {year} {2016})\ p.\ \bibinfo {pages} {WEPMW003}\BibitemShut
  {NoStop}%
\end{thebibliography}%

\end{document}